\documentclass[12pt,a4paper,final]{iopart}

\usepackage{iopams}
\usepackage{graphicx}
\usepackage[colorlinks=true,linkcolor=blue,urlcolor=blue,citecolor=blue]{hyperref}
\usepackage[capitalize,nameinlink,noabbrev]{cleveref}
\usepackage[utf8]{inputenc}
\usepackage[english]{babel}
\usepackage{enumitem}
\usepackage[toc,page]{appendix}
\usepackage{bm}
\usepackage{array}
\usepackage{mdframed}
\usepackage{braket}
\renewcommand\bra[1]{{\langle{#1}|}}
\renewcommand\ket[1]{{|{#1}\rangle}}

\usepackage{comment}
\usepackage{multirow}
\usepackage[format=hang,font=small,labelfont=bf,textfont=sf,figurename=Fig.]{caption}[2013/02/03]
\usepackage[margin=0.2cm]{subcaption}
\usepackage[usenames,dvipsnames,svgnames,table]{xcolor}
\usepackage{breqn}
\usepackage{cite}

\newtheorem{thm}{Theorem}
\newtheorem{lemma}{Lemma}
\newtheorem{proposition}{Proposition}

\newtheorem{defn}{Definition}

\newcommand{\norm}[1]{\left\lVert#1\right\rVert}
\eqnobysec

\begin{document}
	\renewcommand{\equationautorefname}{Eq.} 
	\renewcommand{\figureautorefname}{figure} 
	\renewcommand{\chapterautorefname}{Ch.} 
	\renewcommand{\sectionautorefname}{Sec.} 
	\renewcommand{\subsectionautorefname}{Subsec.} 
	\newcommand{\Mod}[1]{\ (\mathrm{mod}\ #1)}

\title{Finite-key effects in multi-partite quantum key distribution protocols}

\author{Federico Grasselli, Hermann Kampermann and Dagmar Bru\ss}
\address{Institut für Theoretische Physik III, Heinrich-Heine-Universität Düsseldorf, Universitätsstraße 1, D-40225 Düsseldorf, Germany}
\ead{federico.grasselli@hhu.de} 
\begin{abstract}
We analyze the security of two multipartite quantum key distribution (QKD) protocols, specifically we introduce an $N$-partite version of the BB84 protocol and we discuss 
the $N$-partite six-state protocol proposed in \cite{Epping}. 
The security analysis proceeds from the generalization of known results in bipartite QKD to the multipartite scenario, and takes into account finite resources.
In this context we derive a computable expression for the achievable key rate of both protocols by employing the
best-known strategies: the uncertainty relation and the postselection technique.
We compare the performances of the two protocols both for finite resources and infinitely many signals.
\end{abstract}
\vspace*{-3.0ex}
 \bigskip
Quantum Key Distribution (QKD) represents one of the primary applications of quantum information science. Since the proposal of the first QKD protocols \cite{BB84,Ekert},
major advancements have been achieved both on the theoretical and experimental side \cite{ScaraniPasquinucci,DiamantiReview}.
A QKD protocol provides a systematic procedure through which two honest parties 
(Alice and Bob) generate a secret shared key, when connected by an insecure quantum channel and an authenticated insecure classical channel. \\
Recently the generalization of such protocols to multipartite schemes has been investigated \cite{Epping, DICKA}. It has been 
shown that there are quantum-network configurations \cite{Epping} or noise regimes \cite{DICKA} in which the execution of a multipartite scheme is advantageous with
respect to establishing a multipartite secret key via many independent bipartite protocols.
However, the analysis of multipartite QKD protocols has only been carried out in the unrealistic scenario of infinitely many signals exchanged 
through the quantum channel. \\
We compare the performances of two multipartite QKD protocols, which constitute the $N$-partite versions of the asymmetric BB84 \cite{BB84} 
and the asymmetric six-state protocol \cite{Bruss}, and will be denoted as $N$-BB84 and $N$-six-state protocol. 
While the $N$-six-state protocol was first proposed in \cite{Epping}, the $N$-BB84 constitutes a novel multipartite QKD protocol.\\
Our analysis is conducted in the practical case of a finite amount of resources (signals) at the $N$ parties' disposal. The action of a potential eavesdropper (Eve)
on the insecure quantum channel is not restricted at all, as she is allowed to perform any kind of attack (coherent attacks) on the exchanged signals. What is assumed is
that the parties have access to true randomness and that the devices performing measurements on the quantum systems work according to their ideal functionality.\\
The article is structured as follows. In \autoref{MultipartiteQKD} we extend notions and results of bipartite-QKD security analysis to the multipartite scenario. In 
\autoref{NQKD-protocols} we review the $N$-six-state protocol and introduce the $N$-BB84 protocol. Then we obtain a computable expression for their secret key lengths 
in the case of finite resources.
In \autoref{performance} we compare the achievable key rates of the two NQKD protocols in the presence of finite and infinite resources. We conclude the article in 
\autoref{conclusion}.

\section{Multipartite QKD: general framework and achievable key length}  \label{MultipartiteQKD}
Throughout the article we refer to the parties involved in an $N$-partite QKD protocol (NQKD) in the following way: $A$ for Alice, $\textbf{B}$ 
for the set of $N-1$ Bobs, $B_i$ for Bob in position $i$ and $E$ for the eavesdropper Eve. The definitions of distance and entropic quantities employed in this Section
are given in \ref{notation}.\\
The aim of an NQKD protocol is to establish a common secret key, sometimes also referred to as conference key, between all $N$ (trusted) parties.
We consider the following general NQKD protocol. Although the protocol is presented in an entanglement-based view for clarity, there exists an equivalent
prepare-and-measure scheme which requires the adoption of multipartite entangled states only for a small fraction of rounds (see the protocols in
\autoref{NQKD-protocols}). \\
The protocol starts with the distribution of a finite number of signals -described by
genuinely multipartite entangled states- over the insecure quantum channel. All parties perform local measurements on their respective quantum systems, collecting
classical data. A short pre-shared random key indicates to the parties the type of measurement to be performed on each individual state they hold 
(more on this in \autoref{NQKD-protocols}).\\
In the parameter estimation (PE) step the parties reveal a random sample of the collected data, over the insecure classical channel. This allows them to estimate the noise
occurring in the quantum channel and thus to determine the secret key length. At this point the raw keys held by the parties are partially correlated and partially
secret. In order to correct the errors in the raw keys, $A$ performs pairwise an information reconciliation procedure with every $B_i$. 
The procedure consists in some classical communication occurring between $A$ and $B_i$, which allows $B_i$ to compute a guess of $A$'s
raw key. We will refer to this procedure as error correction (EC). 
At last the shared raw key is turned into a secret key with privacy amplification (PA). Each party applies the same randomly chosen hash function
to his/her raw key, where the final length of the key depends on the error rates observed in PE and the desired level of security.
Finally all parties share the same secret key. \\  
During the execution of the NQKD protocol, one or more of the described subprotocols might fail to produce the desired output, thus causing the abortion of the entire 
protocol. In the security analysis this is accounted for by the definition of robustness:
\begin{defn} \label{def-robust}
	{\normalfont \cite{RennerThesis}}. An NQKD protocol is {\normalfont $\varepsilon_{\mathrm{rob}}$-robust} on $\rho_{A\mathbf{B}}$ if, for inputs defined by
	$\rho_{A\mathbf{B}}$, the probability that the protocol aborts is at most $\varepsilon_{\mathrm{rob}}$. 
\end{defn}

In order to study the effects of finite resources on an NQKD protocol, one needs to extend the concept of $\varepsilon$-security of a key
\cite{RennerThesis} to the multi-partite scenario:
\begin{defn} \label{def-secure}
	{\normalfont \cite{RennerThesis},\cite{Portman}.} Let $\rho_{A\mathbf{B}E}$ be a density operator. Any NQKD protocol, which is $\varepsilon_{\mathrm{rob}}$-robust
	on $\Tr_E[\rho_{A\mathbf{B}E}]$, is said to be {\normalfont $\varepsilon_{\mathrm{tot}}$-secure} on $\rho_{A\mathbf{B}E}$ if the following inequality holds:
	\begin{equation}
	(1-\varepsilon_{\mathrm{rob}}) \frac{1}{2} \norm{\rho_{S_A \mathbf{S_B}E'}-\rho_{\mathbf{U}}\otimes \rho_{E'}} \leq \varepsilon_{\mathrm{tot}}
	\end{equation}
	where $\rho_{S_A\mathbf{S_B}E'}$ is the density operator describing the final keys held by the $N$ parties and Eve's enlarged subsystem 
	$\mathcal{H}_{E'}$ (including the information of the classical channels), 
	while $\rho_{\mathbf{U}}$ is the uniform state on the key space of the $N$ parties:
	\begin{equation}
	\rho_{\mathbf{U}} \equiv \sum_{s\in \mathcal{S}}\frac{1}{|\mathcal{S}|}\bigotimes_{i=1}^{N} \ket{s}\bra{s} 
	\end{equation}
	with $\mathcal{S}$ the set of possible secret keys.
\end{defn}
The total security parameter $\varepsilon_{\mathrm{tot}}$ quantifies the deviation of the NQKD protocol from an ideal protocol, i.e. one that either outputs a set of
perfectly-correlated and fully-secret keys or aborts. In other words, an NQKD protocol is $\varepsilon_{\mathrm{tot}}$-secure if it behaves like an ideal protocol 
except for probability $\varepsilon_{\mathrm{tot}}$. With this definition, the parameter that actually accounts for the correctness and secrecy of the protocol when it
does not abort, is: $\varepsilon_{\mathrm{tot}}/(1-\varepsilon_{\mathrm{rob}})$. An NQKD protocol may deviate from an ideal one if, for instance, its EC procedure fails
to correct all the errors between $A$ and \textbf{B}'s strings. In particular, if the probability that at least one $B_i$ holds a different string than $A$ -after EC- is
$\varepsilon_{\mathrm{EC}}$, then the NQKD protocol is $\varepsilon_{\mathrm{tot}}$-secure, with $\varepsilon_{\mathrm{tot}}\geq \varepsilon_{\mathrm{EC}}$.
Formally, the EC failure probability is defined as:

\begin{defn} \label{def-IRsecurity}
	{\normalfont  \cite{RennerThesis}}. 
	Let $P_{X\mathbf{K}}$ be a probability distribution. Any set of error correction protocols
	$\{\mathsf{EC}_i\}_{i=1}^{N-1}$, which is $\varepsilon_{\mathrm{rob}}$-robust on $P_{X\mathbf{K}}$, is said to be {\normalfont $\varepsilon_{\mathrm{EC}}$-secure} on
	$P_{X\mathbf{K}}$ if the following holds:
	\begin{equation}
	(1-\varepsilon_{\mathrm{rob}}) \mathrm{Pr}\left[\exists i\in\{1,\dots,N-1\} : \hat{k}_i \neq x\right] \leq \varepsilon_{\mathrm{EC}}  \nonumber  
	\end{equation}
	where the guess $\hat{k}_i$ is computed by $B_i$ according to protocol $\mathsf{EC}_i$, and the probability is computed for inputs $(x,\mathbf{k})$ chosen according 
	to $P_{X\mathbf{K}}$, conditioned on the fact that no $\mathsf{EC}_i$ aborted. If $\{\mathsf{EC}_i\}_{i=1}^{N-1}$ is $\varepsilon_{\mathrm{EC}}$-secure for any
	probability distribution, it is $\varepsilon_{\mathrm{EC}}$-fully secure.
\end{defn}

In this article we assume that the NQKD protocol may abort only during the EC procedure. Thus the abortion probability of the chosen set of EC procedures is also the
abortion probability of the whole protocol\footnote[3]{Note, however, that a higher global abortion probability for fixed security parameter 
$\varepsilon_{\mathrm{tot}}$ may lead to higher key rates.}. \\
The classical communication occurring during EC contains some information about the key. The amount of information about the key that is leaked to $E$ from the insecure
classical channel is quantified by the leakage:
\begin{defn} \label{def-leak}
	{\normalfont  \cite{RennerThesis}}.
	Let $\{\mathsf{EC}_i\}_{i=1}^{N-1}$ be a set of EC protocols. The NQKD protocol adopting such a set of protocols for error correction has {\normalfont leakage}:
	\begin{equation}
	\mathrm{leak}_{\{\mathsf{EC}_i\}}^{\mathrm{NQKD}} \equiv \log_2 \lvert \mathcal{C}_{1,\dots,N-1} \rvert - 
	\min_{x,\mathbf{k}} H_{\mathrm{min}}\left(P_{\mathbf{C}|X=x,\mathbf{K}=\mathbf{k}}\right)   \nonumber
	\end{equation}
	where $\mathcal{C}_{1,\dots,N-1}$ is the set of $(N-1)$-tuples representing all possible communication transcripts allowed by the chosen EC protocols, i.e.:
	\begin{equation}
	\mathcal{C}_{1,\dots,N-1}=\left\lbrace (c_1,\dots,c_{N-1}) : P_{\mathbf{C}}(c_1,\dots,c_{N-1})\neq 0 \right\rbrace
	\,\,,\label{setofcommunications}
	\end{equation}
	$P_{\mathbf{C}|X=x,\mathbf{K}=\mathbf{k}}$ is the transcripts' distribution conditioned on $A$ and $\mathbf{B}$'s raw keys and
	$H_{\mathrm{min}}\left(P_{\mathbf{C}|X=x,\mathbf{K}=\mathbf{k}}\right)$ is the min-entropy defined on a probability distribution
	(\ref{min-entropy-trivial},\ref{min-entropy-probability}).
\end{defn}

We now present our results on the achievable key length (\autoref{achievable1and2way}) and the minimum leakage (\autoref{leak_upperbound}) of a general 
$\varepsilon_{\mathrm{tot}}$-secure NQKD protocol, which constitute a generalization of analogous results \cite[Lemmas 6.4.1 and 6.3.4]{RennerThesis} valid for 
bipartite QKD. The general structure of the proofs is derived from the bipartite case, but deals with the new definitions of security and leakage (\autoref{def-secure},
\ref{def-IRsecurity}, \ref{def-leak}) for multipartite schemes. As in the bipartite case, the security of an NQKD protocol can be inferred by correctness and 
secrecy (\ref{NQKD_def_and_proofs}). While the correctness of a protocol is determined by its EC procedure, the secrecy is linked to the final-key length via the 
leftover hashing lemma \cite[Corollary 5.6.1]{RennerThesis}.
In fact, in PA the parties map their shared key to another key which is short enough to be secret (i.e. unknown to the eavesdropper Eve). In \autoref{achievable1and2way}
we present the achievable key length of an $\varepsilon_{\mathrm{tot}}$-secure NQKD protocol for a general two-way EC procedure, while typically only the special case of
one-way EC is addressed. This is achieved thanks to the result on the information leakage with two-way EC presented in \ref{RennerThm} \cite{privateComm}. 
A detailed version of the proofs of \autoref{achievable1and2way} and \autoref{leak_upperbound} is presented in \ref{NQKD_def_and_proofs}.
\begin{thm} \label{achievable1and2way}
	Let: $\bar{\varepsilon}> 0$, $\varepsilon_{\mathrm{EC}}> 0$, $\varepsilon_{\mathrm{PA}}>0$, $\varepsilon_{\mathrm{rob}}\geq 0$
	and $\rho_{A\mathbf{B}E}$ be a density operator.
	Let $\rho_{X\mathbf{K}E}$ be the output -prior to EC and PA- of an NQKD protocol applied to $\rho_{A\mathbf{B}E}$. If the two-way EC protocol
	$\{\mathsf{EC}_i\}_{i=1}^{N-1}$ is $\varepsilon_{\mathrm{EC}}$-secure and $\varepsilon_{\mathrm{rob}}$-robust
	on the distribution defined by $\rho_{X\mathbf{K}}$, and if 
	$\mathsf{PP}_{\{\mathsf{EC}_i\},\mathcal{F}}$ is the post-processing protocol defined by the set of EC protocols and by the set 
	of two-universal hash functions $\mathcal{F}$ with co-domain $\{0,1\}^{\ell}$ such that\footnote[7]{The $\bar{\varepsilon}$-environment of the min-entropy is defined 
	via the purified distance, see \ref{notation}.} the secret key length $\ell$ fulfills:
	\begin{equation}
	\ell \leq H_{\mathrm{min}}^{\bar{\varepsilon},\,\mathrm{P}} \left(\rho_{XE}\rvert E\right) - \mathrm{leak}_{\{\mathsf{EC}_i\}}^{\mathrm{NQKD}}
	- 2\log_2 \frac{1-\varepsilon_{\mathrm{rob}}}{2\,\varepsilon_{\mathrm{PA}}}  \,\,, \label{upperbound-2wayIR}
	\end{equation}   
	then the NQKD protocol is $\varepsilon_{\mathrm{tot}}$-secure on $\rho_{A\mathbf{B}E}$, where $\varepsilon_{\mathrm{tot}}$ is defined as:
	\mbox{$\varepsilon_{\mathrm{tot}}=2\bar{\varepsilon}+\varepsilon_{\mathrm{EC}}+\varepsilon_{\mathrm{PA}}$.} \\
	If one restricts to one-way EC, the same result holds but with the $\bar{\varepsilon}$-environment of the min-entropy defined via the trace distance.
\end{thm}
\begin{thm} \label{leak_upperbound}
	Given a probability distribution $P_{X\mathbf{K}}$, there exists a 1-way EC protocol that is: $\varepsilon_{\mathrm{EC}}$-fully secure, $2(N-1)\varepsilon'$-robust
	on $P_{X\mathbf{K}}$, and has leakage:
	\begin{equation}
	\mathrm{leak}_{\mathrm{EC}}^{\mathrm{NQKD}} \leq \max_i H_{0}^{\varepsilon'}(P_{XK_i} \rvert K_i) + 
	\log_2 \frac{2(N-1)}{\varepsilon_{\mathrm{EC}}}  \label{leak-upperbound}
	\end{equation}
\end{thm}
The upper bound in \autoref{leak_upperbound} is independent of the EC protocol, 
thus also bounds the leakage of an \emph{optimal} 1-way EC protocol which is $\varepsilon_{\mathrm{EC}}$-fully secure and \mbox{$2(N-1)\varepsilon'$-robust} on
$P_{X\mathbf{K}}$.

\section{N-BB84 and N-six-state protocol}  \label{NQKD-protocols}
Here we present the two NQKD protocols whose performance will be investigated in \autoref{performance}.
We introduce the $N$-BB84 protocol which is the $N$-partite version of the asymmetric BB84 protocol \cite{BB84}:
\begin{mdframed}
	\begin{center}
		\textbf{\large $N$-BB84 protocol}
	\end{center}
\begin{enumerate}
	\item Distribution of $N$-qubit GHZ states:
		\begin{equation}
		\ket{\mathrm{GHZ}}_N \equiv \frac{1}{\sqrt{2}} \left(\ket{0}^{\otimes N} + \ket{1}^{\otimes N}\right)   \label{GHZ}
		\end{equation} for $L$ rounds.
	\item In 1st-type rounds each party measures in the $Z$-basis, in 2nd-type rounds -which occur with probability $p$\footnote[9]{$ L \cdot h(p)$ bits of preshared
		secure key are used to mark the 2nd-type rounds.}- each party measures in the $X$-basis. The total number of 2nd-type rounds is: $m=Lp$.
	\item Parameter estimation:
	\begin{enumerate}
		\item Computation of $Q_{A B_i}^m=(1-\braket{Z_A Z_{B_i}}_m)/2$ for every $B_i$, where $Z_A Z_{B_i}$ is averaged over $m$ 1st-type rounds randomly chosen by Alice.
		In the ideal situation: $Q_{A B_i}^m=0$.
		\item Computation of $Q_X^m= (1-\braket{X^{\otimes N}}_m)/2$, where $X^{\otimes N}$ is averaged over the 2nd-type rounds.
		Note that in the ideal situation: $Q_X^m=0$ \cite{Epping}.
	\end{enumerate}
	\item The secret key is obtained from the remaining data of $n=L-2m$ 1st-type rounds. 
	\item Classical post-processing: 
	\begin{enumerate}
		\item $A$ sends the same EC information to every $B_i$.
		\item $A$ and \textbf{B} apply the same two-universal hash function to their corrected data.
	\end{enumerate}
\end{enumerate}
\end{mdframed}
\textit{Remarks}: Note that the frequencies $Q_{A B_i}^m$ and $Q_X^m$ observed in the PE step are the fraction of discordant $Z$-outcomes between $A$ and $B_i$ and 
the frequency of the outcome $-1$ when the parties measure the operator $X^{\otimes N}$, respectively.\\
In an equivalent prepare-and-measure scheme, Alice directly produces the $(N-1)$-qubit projection of the GHZ state according to her fictitious random outcome and
distributes it to the Bobs. In particular, she prepares product states if the $Z$-basis is chosen and multipartite entangled states when the $X$-basis is picked. Thus 
the production of multipartite entangled states is only required for $Lp$ rounds, while in all other rounds product states are prepared \cite{Epping}.\\
For the protocol's security to hold, the preshared secret key indicating to the parties the 2nd-type rounds needs to be refreshed at every new execution 
of the protocol. Therefore, the net amount of new secret key bits produced by one run of the protocol is obtained by subtracting $L \cdot h(p)$ bits from the final
key length presented in \autoref{computablerates}. We take into account this term for both protocols when investigating their performance in
\autoref{performance}.\medskip\\
We refer to \cite{Epping} for a detailed description of the steps characterizing the $N$-six-state protocol. However, the only actual differences with respect to the 
$N$-BB84 protocol are that: in the 2nd-type rounds each party measures randomly in the $X$- or $Y$-basis and all parties jointly flip 
their $Z$-measurement outcomes with probability 1/2. The bits to be flipped can be announced by Alice after the distribution and measurement of the states.
These operations enable the implementation of the extended depolarization procedure \cite{Epping} on the classical
data, without adding further quantum gates.\\
The frequencies observed in the PE step of the $N$-six-state protocol are again $Q_{A B_i}^m$ and $Q_X^{m'}$\footnote[1]{Since the value of $X^{\otimes N}$ must be
registered only when an even number of parties measured in the $Y$ basis, $m'=m/2$. See \cite{Epping} for further details.}, 
plus $Q^m_Z$, i.e. the fraction of rounds in which at least one Bob measured a different $Z$-outcome than $A$'s. We will refer to the corresponding probabilities as:
$P_{AB_i}$, $P_X$ and $P_Z$.\medskip\\
The frequencies observed in the PE steps of both protocols enable to quantify the amount of noise occurring in the quantum channel.
However, these statistics are collected on finite-size samples, thus they only represent an estimate of the channel's noise. In 
\ref{PE} we quantitatively describe how the finite statistics of PE characterize the quantum channel's noise, for both NQKD protocols.

\subsection{Computable key length} \label{computablerates}
In order to employ the results of \autoref{MultipartiteQKD} in a performance comparison of the two NQKD protocols one needs to characterize $E$'s knowledge about the
key. This is achieved by assigning the noise in the quantum channel to eavesdropping.
This means, in practice, that one can bound the unknown entropies with quantities exclusively depending on the noise affecting the
quantum channel. In turn, the channel's noise is characterized by the finite PE statistics, as explained above. \\
As a result, we obtain a computable expression for the achievable key length of both protocols, that is an expression solely depending on the observed PE statistics,
the desired level of security, and the total number of quantum signals.\\
The techniques we adopt to obtain a computable key length are the following. We employ the uncertainty relation (for smooth entropies) presented in \cite{uncertainty-rel}
for the $N$-BB84 protocol, thus showing its first application to NQKD. 
For the $N$-six-state protocol we instead employ the Postselection technique (PS) \cite{postsel} in combination with the Asymptotic
Equipartition Property (AEP) \cite{RennerThesis}, and we exploit the symmetries induced by the extended depolarization procedure.\\
We arrive at the computable key lengths of the $N$-BB84 and $N$-six-state protocol:
\begin{thm}  \label{computable-length-NBB84}
	The $N$-BB84 protocol, with the optimal 1-way EC protocol (which is $\varepsilon_{\mathrm{EC}}$-fully secure and
	$2(N-1)\varepsilon_{\mathrm{PE}}\,$-robust) and where the secret key generated by two-universal hashing has length
	\begin{eqnarray}
	\fl \ell= n\, \left[1 - h\left(Q^m_X + 2\xi(\varepsilon_x,n,m)\right) - \max_i h\left(Q_{AB_i}^m +2\xi(\varepsilon_z,n,m)\right)\right] 
	- \log_2 \frac{2(N-1)}{\varepsilon_{\mathrm{EC}}}  \nonumber \\
	\fl - 2\log_2 \frac{1-2(N-1)\varepsilon_{\mathrm{PE}}}{2\,\varepsilon_{\mathrm{PA}}}  \quad, \label{computable-length-NBB84-expression}
	\end{eqnarray}
	is $\varepsilon_{\mathrm{tot}}$-secure with $\varepsilon_{\mathrm{tot}}=2\varepsilon_{\mathrm{PE}} + \varepsilon_{\mathrm{EC}} + \varepsilon_{\mathrm{PA}}$,
	where $\varepsilon_{\mathrm{PE}}$ is defined as (\ref{epsilon-PE-2}):
		\begin{equation}
		\varepsilon_{\mathrm{PE}} \equiv \sqrt{(N-1)\varepsilon_z + \varepsilon_x} 
		\end{equation} 
		and $\xi(\varepsilon,n,m)$ as (\ref{Th-xi}):
		\begin{equation}
		\xi(\varepsilon,n,m) \equiv \sqrt{\frac{(n+m)(m+1)}{8nm^2} \ln \left(\frac{1}{\varepsilon}\right)}  \,\,. 
		\end{equation}
\end{thm}
\begin{thm}  \label{computable-length-Nsixstate}
	The $N$-six-state protocol, with the optimal 1-way EC protocol (which is $\varepsilon_{\mathrm{EC}}$-fully secure and
	$2(N-1)\varepsilon_{\mathrm{PE}}\,$-robust) and where the secret key generated by two-universal hashing has length
	\begin{eqnarray}\fl
	\ell= \,n \, \inf_{\Gamma_{\mathrm{PE}}} 
	\left[\left(1-\frac{P_Z}{2} -P_X\right)\log_2 \left(1-\frac{P_Z}{2} -P_X\right) + \left(P_X -\frac{P_Z}{2}\right) \log_2 \left(P_X -\frac{P_Z}{2}\right) \right. 
	\nonumber \\ 
	\fl \left. + (1-P_Z) \left(1-\log_2 (1-P_Z)\right)- 5\sqrt{\frac{\log_2 (1/\bar{\varepsilon})}{n}} - \max_i h\left(P_{AB_i}\right) - \log_2 (5) \, 
	\sqrt{\frac{2\log_2 (1/(2\varepsilon_{\mathrm{PE}}))}{n}}\right]  \nonumber \\
	\fl - \log_2 \frac{2(N-1)}{\varepsilon_{\mathrm{EC}}}  - 2\log_2 \frac{1-2(N-1)\varepsilon_{\mathrm{PE}}}{2\,\varepsilon_{\mathrm{PA}}} -2(2^{2N}-1)\log_2 (L+1) 
	\quad,	\label{computable-length-Nsixstate-expression}
	\end{eqnarray}
	is $\varepsilon_{\mathrm{tot}}$-secure with $\varepsilon_{\mathrm{tot}}=(L+1)^{(2^{2N}-1)}(2\bar{\varepsilon} + \varepsilon_{\mathrm{PE}} + \varepsilon_{\mathrm{EC}}
	+ \varepsilon_{\mathrm{PA}})$,	where $P_X$, $P_{A B_i}$ and $P_{Z}$ are minimized over the set:
	\begin{eqnarray}
	 \Gamma_{\mathrm{PE}} &\equiv \left\lbrace P_{AB_i}, P_Z, P_X : \frac{1}{2} |Q_{AB_i}^m - P_{AB_i}| \leq \eta(\varepsilon_z,2,m)\,\,\forall\, i \right. 
	 \nonumber \\
	 &\left.\wedge\, \frac{1}{2} |Q^{m'}_X - P_X| \leq \eta(\varepsilon_x,2,m')\,\wedge\, \frac{1}{2} |Q_{Z}^m - P_Z| \leq \eta(\varepsilon_z',2,m) \right\rbrace\,\,.
	 	\label{GammaPE}
	\end{eqnarray}
	The parameters $\varepsilon_x,\varepsilon_z,\varepsilon_z'$ are linked to $\varepsilon_{\mathrm{PE}}$ via (\ref{epsilon-PE-1}):
	\begin{equation}
		\varepsilon_{\mathrm{PE}} \equiv \varepsilon_z' + (N-1)\varepsilon_z +\varepsilon_x  
	\end{equation}
	while $\eta(\varepsilon,d,m)$ is defined as (\ref{eta}):
	\begin{equation}
		\eta(\varepsilon,d,m) \equiv \sqrt{\frac{\ln (1/\varepsilon)+d\ln (m+1)}{8 m}} \quad.
	\end{equation}
\end{thm}
For the derivation of \autoref{computable-length-NBB84} and \autoref{computable-length-Nsixstate}, we refer to \ref{details-computablerates}.

\section{Performance comparison} \label{performance}
We compare the performances of the two NQKD protocols by studying their secret-key rates, i.e. the fraction of shared secret bits per transmitted quantum signal
($\ell/L$).
For this purpose we investigate the computable key lengths (\ref{computable-length-NBB84-expression}) and (\ref{computable-length-Nsixstate-expression})-
corrected with the term ``$-L\cdot h(p)$'' that accounts for the preshared secret key- for a given
number of parties $N$ and a fixed total security parameter $\varepsilon_{\mathrm{tot}}$.\\
In order to carry out a fair comparison, we assume that the PE statistics of both protocols are generated by the same error model.

\subsection{Error model} \label{errormodel}
We assume that in every distribution round white noise acted on the ideal state and that the action of the noise is the same
in every round\footnote{The same error model is used, for instance, in \cite{Epping}.}.
The total distributed state over all rounds is a product state: $\rho^{\otimes L}_{A\mathbf{B}}$, where the single-round state is given by:
\begin{eqnarray}
\rho_{A\mathbf{B}} = (1-\nu)\ket{\mathrm{GHZ}}_N\bra{\mathrm{GHZ}}_N +\nu\, \frac{\mathrm{id}_{A\mathbf{B}}}{2^N}  \label{mixed-whitenoise}
\end{eqnarray}
where $\nu$ is the noise parameter and $\ket{\mathrm{GHZ}}_N$ is the GHZ state of $N$ qubits (\ref{GHZ}).\\
The state (\ref{mixed-whitenoise}) can be seen as the result of the action of a depolarizing channel on the whole $N$-qubit system, such that it is diagonal 
in the GHZ basis \cite{Epping} and the probabilities $P_{AB_i}$ (of $A$ and $B_i$ having discordant $Z$-outcomes), $P_X$ 
(of having the outcome $-1$ when the parties measured $X^{\otimes N}$) and $P_Z$ (of having at least one Bob with a different 
$Z$-outcome than $A$'s) are given by:
\begin{eqnarray}
P_{AB_i} &= \nu / 2  \quad\forall\, i \label{PABi-noise}  \\
P_X &= P_{AB}  \label{PX-noise} \\
P_Z &= \frac{2^N -2}{2^{N-1}} P_{AB}  \label{PZ-noise} \,\,.
\end{eqnarray}
For ease of notation we will drop the index $i$ in the probabilities $P_{AB_i}$.
We assume that the frequencies $Q^m_{AB_i}$, $Q^m_X$ and $Q^m_Z$ observed in the PE step of both protocols are linked by the same relations 
(\ref{PX-noise}), (\ref{PZ-noise}) that hold for the corresponding probabilities.

\subsection{Infinite resources} \label{infiniteresources}
\begin{figure}[!htb]
	\centering
	\includegraphics[width=0.7\linewidth,keepaspectratio]{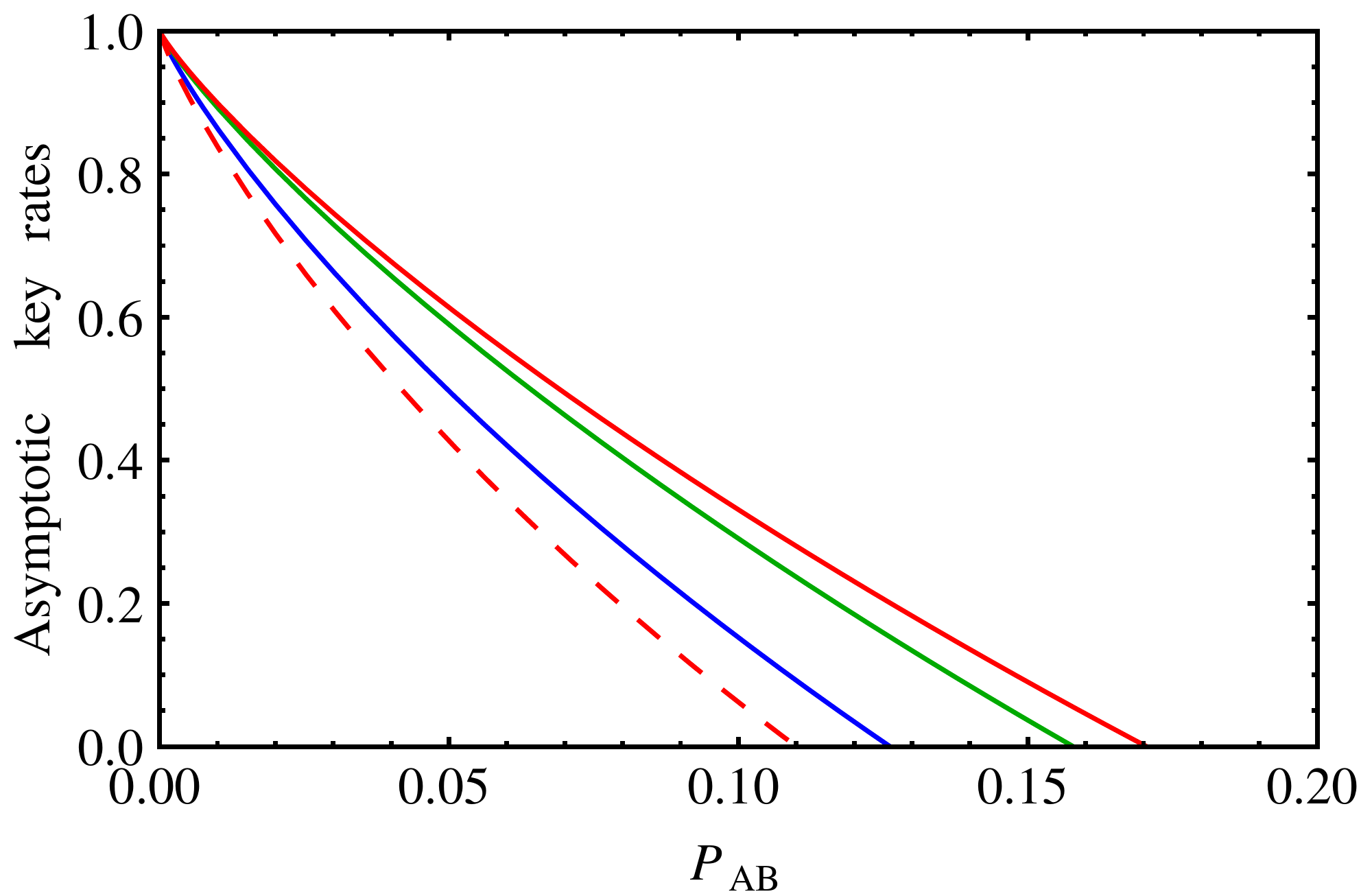}
	\caption{Asymptotic key rates ($N$-six-state solid, $N$-BB84 dashed) for $N=2,5,\infty$ (blue, green, red) as a function of the probability of
	discordant $Z$-outcomes between $A$ and $B_i$ ($P_{AB}$), in the presence of a global depolarizing channel (\ref{mixed-whitenoise}).
	Due to the symmetric action of the white noise on the quantum channel: $P_X=P_{AB}$.
	The $N$-BB84 asymptotic key rate presents only one curve since it is independent of $N$.}
	\label{asymptotic_plot1}
\end{figure}
\begin{figure}[!htb]
	\begin{minipage}[c]{0.60\textwidth}
		\centering
		\includegraphics[width=1\linewidth,keepaspectratio]{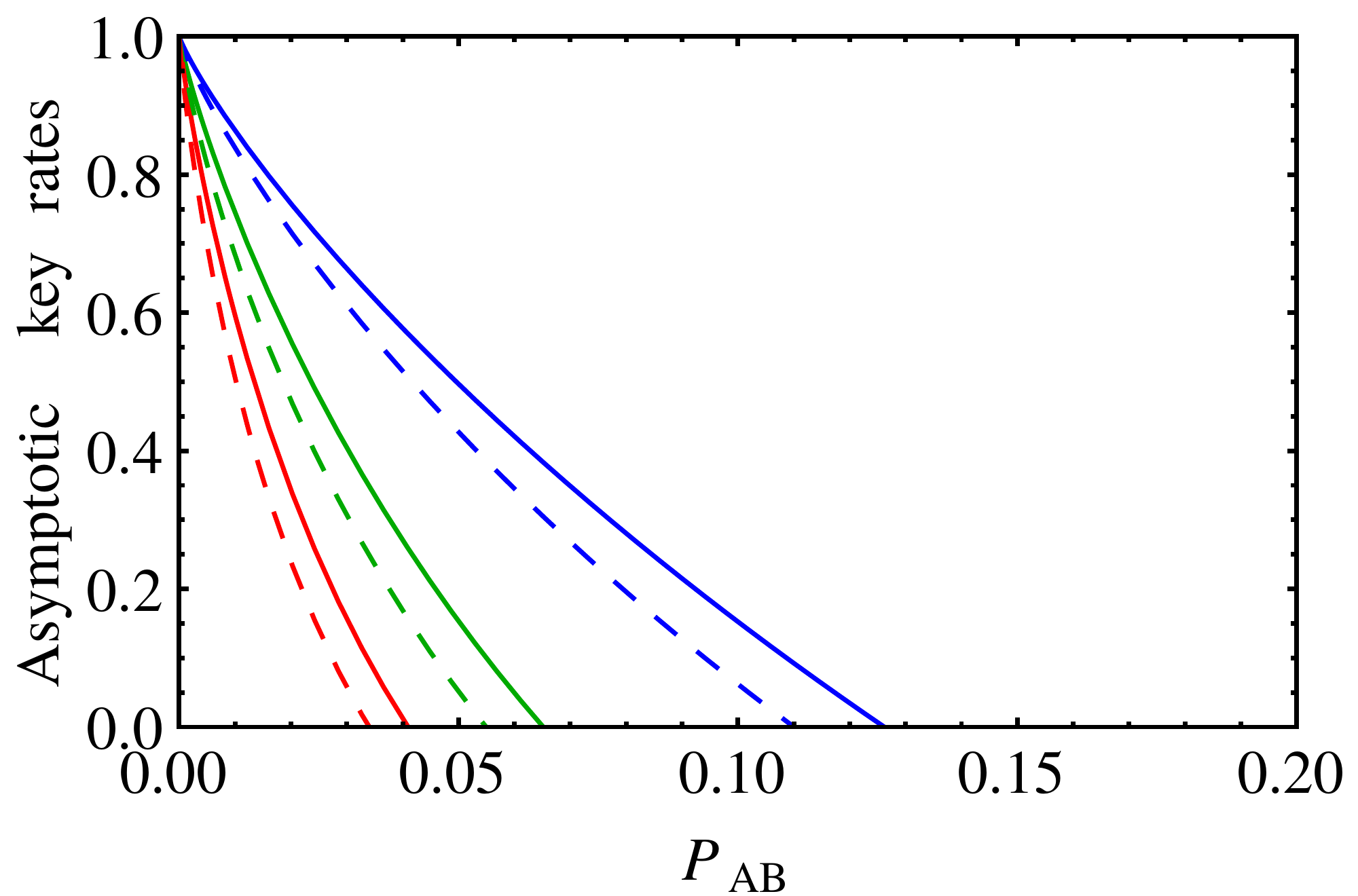}
	\end{minipage}
	\begin{minipage}[c]{0.40\textwidth}
		\caption{Asymptotic key rates ($N$-six-state solid, $N$-BB84 dashed) for $N=2,5,10$ (blue, green, red) as a function of the probability of
		discordant $Z$-outcomes between $A$ and $B_i$ ($P_{AB}$), in the presence of local depolarizing channels (\ref{new-single-round}). 
		With this model the rate of both protocols decreases for increasing number of parties and fixed $P_{AB}$.}
		\label{asymptotic_plot2}
	\end{minipage}
\end{figure}

In the asymptotic limit of infinitely many rounds ($L\rightarrow \infty$), all the correction terms due to finite statistics vanish, as well as all the correction terms
due to the $\varepsilon$-security of the key. For instance, the PE frequencies coincide with their corresponding probabilities. \\
For the assumed error model, the asymptotic key rates of the $N$-six-state protocol ($r_{6\mbox{\footnotesize-state}}$) and the $N$-BB84 protocol 
($r_{\mathrm{BB84}}$) read: 
\begin{eqnarray}
\fl r_{6\mbox{\footnotesize-state}} (P_{AB},N) =   
\left(1-\frac{P_Z}{2} -P_{AB}\right)\log_2 \left(1-\frac{P_Z}{2} -P_{AB}\right) 
\nonumber \\
\fl + \left(P_{AB} -\frac{P_Z}{2}\right) \log_2 \left(P_{AB} -\frac{P_Z}{2}\right) + (1-P_Z) \left(1-\log_2 (1-P_Z)\right) - h\left(P_{AB}\right)
\label{asymptotic-Nsixstate}   \\[0.3cm] 
\fl r_{\mathrm{BB84}} (P_{AB}) = 1-2 h(P_{AB})   \label{asymptotic-NBB84}
\end{eqnarray}
where $P_Z$ is fixed by (\ref{PZ-noise}) and the rates have been maximized over the probability $p$ of performing 2nd-type rounds. 
For $N=2$ the rate (\ref{asymptotic-Nsixstate}) reduces to the asymptotic rate of the bipartite six-state protocol \cite{ScaraniPasquinucci}, while
(\ref{asymptotic-NBB84}) is independent of $N$ -for fixed $P_{AB}$- and coincides with the asymptotic bipartite BB84 rate \cite{ScaraniPasquinucci}.
The reason for which (\ref{asymptotic-NBB84}) does not depend on $N$ is that the $N$-BB84 protocol -unlike the $N$-six-state- does not completely characterize the state
shared by all the parties, thus its asymptotic rate only depends on $P_{AB}$ and $P_X$. For the highly symmetric error model introduced in \autoref{errormodel}, it 
holds: $P_X=P_{AB}=\nu/2$ which is independent of the number of parties involved.\\
In \autoref{asymptotic_plot1} we plot the asymptotic rate of both protocols as a function of the probability of discordant raw key bits between $A$ and $B_i$ ($P_{AB}$),
for various numbers of parties $N$. 
By noting that the $N$-six-state protocol outperforms the $N$-BB84 for equal $P_{AB}$ and any number of parties $N$, we observe in the 
$N$-partite asymptotic scenario that a six-state-type protocol produces higher rates than a BB84 one, extending known results of the 
bipartite case \cite{ScaraniPasquinucci}. \\
Interestingly, the rate of both protocols does not decrease for an increasing number of parties and fixed $P_{AB}$.
However, one should keep in mind that increasing $N$ for fixed $P_{AB}$ may not be physically reasonable. In fact, 
according to our error model, if $P_{AB}$ is fixed then also the noise parameter $\nu$ (quantifying the amount of depolarization on all $N$ qubits) is fixed, and
increasing $N$ with a fixed noise parameter may not describe realistic quantum channels. 
Consider, for instance, the case in which part of the noise generating $P_{AB}$ is due to the failure of imperfect bipartite gates
used for the distribution of the GHZ state. Then an increase of $N$, obtained by adding gates with the same failure probability, would lead to an
increase of $P_{AB}$ \cite{Epping}.\\
Moreover, the adoption of other error models can lead to key rates decreasing in the number of parties, for fixed $P_{AB}$. For instance if the noise on the ideal
distributed state is modeled as the independent action of the depolarizing map
\begin{equation}
	\mathcal{D}(\rho) = (1-\nu) \rho + \nu \,\frac{\mathrm{id_2}}{2}  \label{depolarizing-map}
\end{equation}
on each $B_i$, i.e. the single-round state reads:
\begin{equation}
  \rho_{A\mathbf{B}} = 	\mathcal{D}^{\otimes (N-1)} \left(\ket{\mathrm{GHZ}}_N\bra{\mathrm{GHZ}}_N\right) \,\,,  \label{new-single-round}
\end{equation}
then the probabilities of interest are given by:
\begin{eqnarray}
P_{AB} &= \nu / 2  \label{PABi-noise-new}  \\
P_X &= \frac{1-(1-2P_{AB})^{N-1}}{2}  \label{PX-noise-new} \\
P_Z &= 1-(1-P_{AB})^{N-1}  \label{PZ-noise-new}
\end{eqnarray}
where we dropped the index $i$ in the probabilities $P_{AB_i}$.
The asymptotic key rates of the $N$-BB84 and $N$-six-state protocol computed with the new probabilities (\ref{PABi-noise-new}), (\ref{PX-noise-new})
and (\ref{PZ-noise-new}) decrease for increasing number of parties, see \autoref{asymptotic_plot2}.

\subsection{Finite resources}  \label{finiteresources}
\begin{figure}[!hb]
	\centering
	\begin{subfigure}[t]{.5\textwidth}
		\centering
		\includegraphics[width=1\linewidth,keepaspectratio]{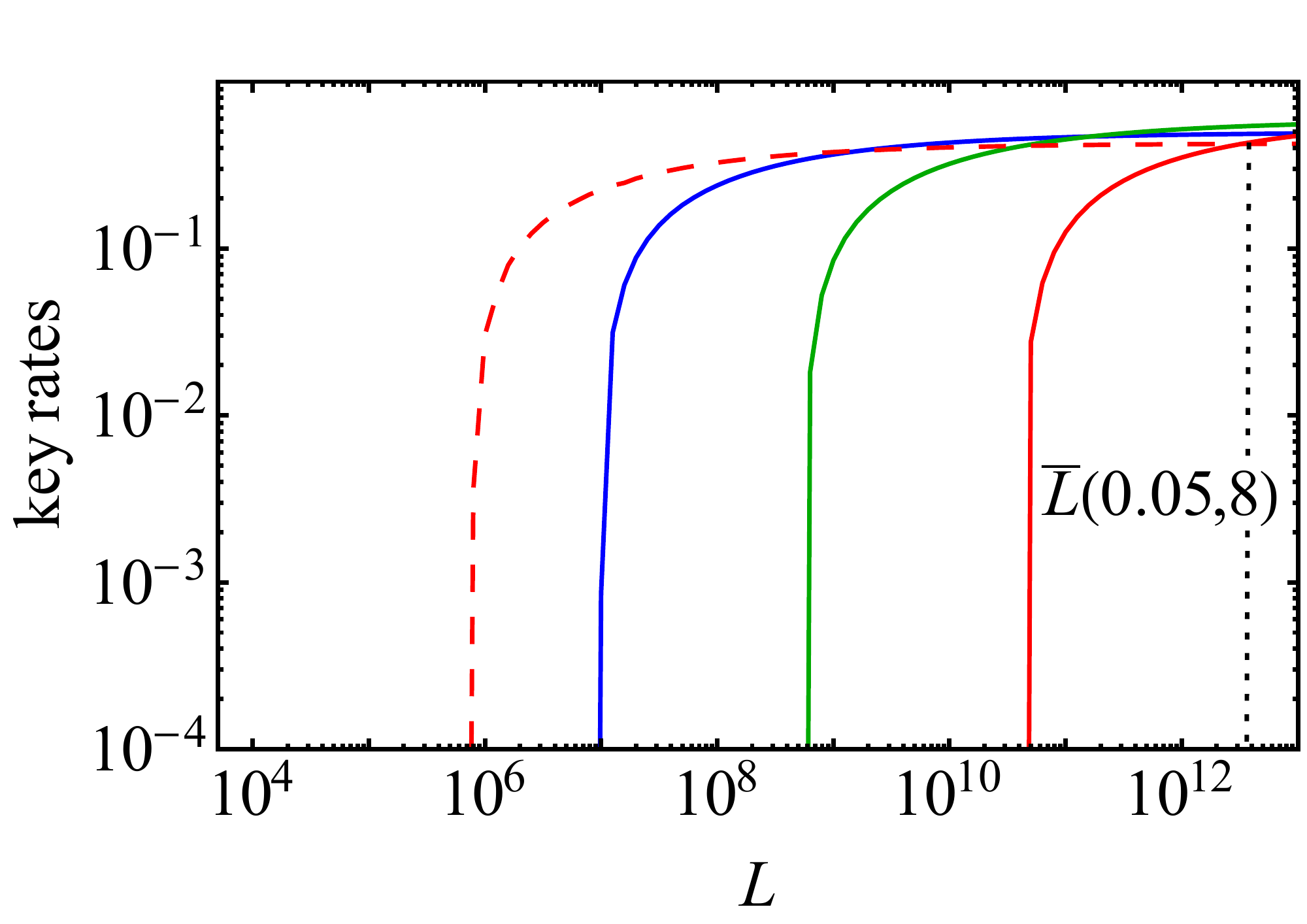}
		\caption{Key rates as a function of the total number of rounds $L$ for \mbox{$N=2,5,8$} (blue, green, red; left to right) and fixed
			\mbox{$Q^m_{AB}=0.05$}. Note that even for finite number of rounds the $N$-BB84 rate is approximately independent of $N$.}
		\label{fixedQincreasingN}
	\end{subfigure}%
	\begin{subfigure}[t]{.5\textwidth}
		\centering
		\includegraphics[width=1\linewidth,keepaspectratio]{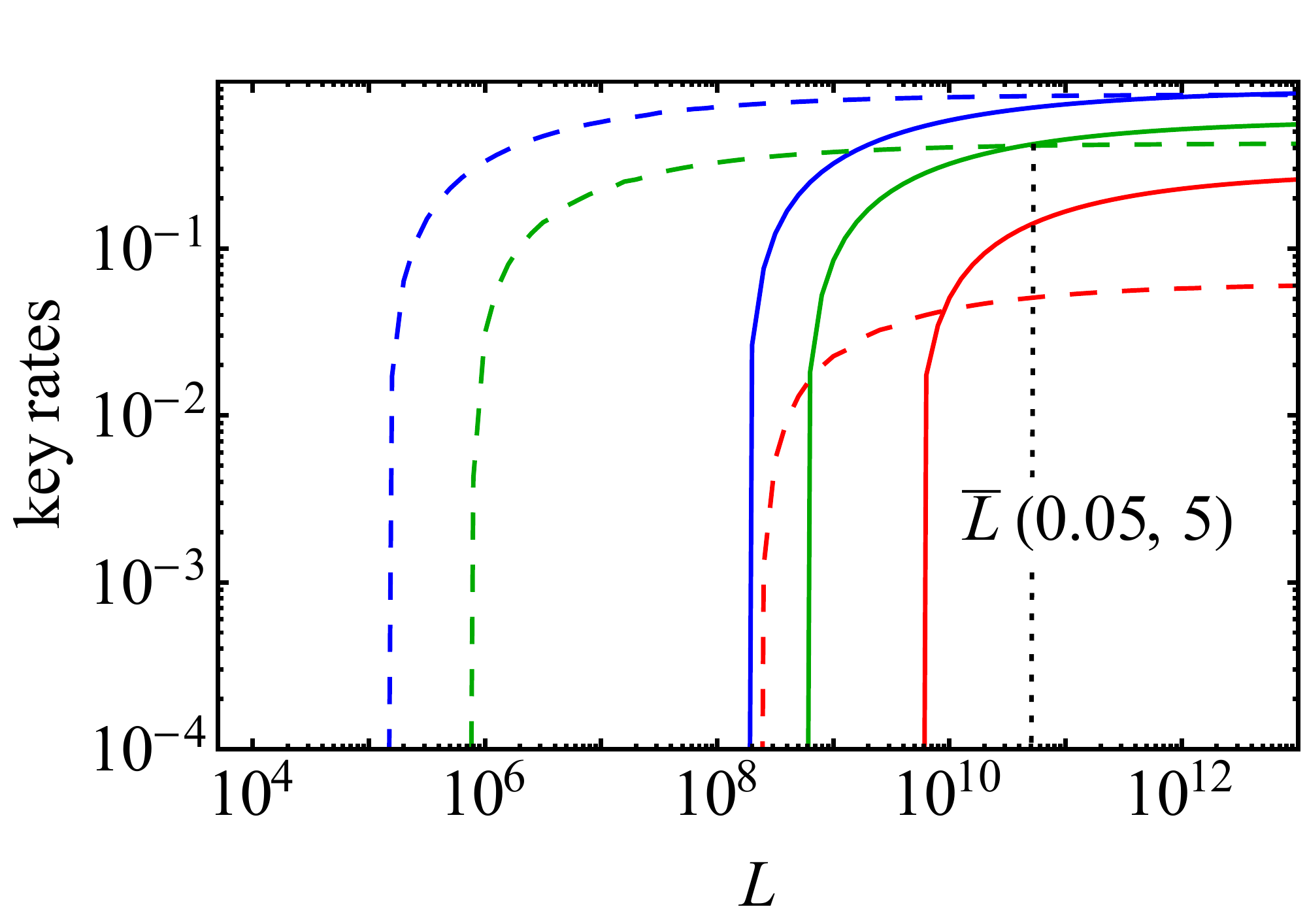}
		\caption{Key rates as a function of the total number of rounds $L$ for \mbox{$Q^m_{AB}=0.01,0.05,0.1$} (blue, green, red; left to right) 
			and fixed $N=5$.}
		\label{fixedNincreasingQ}
	\end{subfigure}
	\caption{Key rates ($N$-six-state solid,$N$-BB84 dashed) as a function of the number of signals $L$.}
	\label{finite-rates}
\end{figure}
\begin{figure}[!htb]
	\centering
	\begin{subfigure}[t]{.5\textwidth}
		\centering
		\includegraphics[width=1\linewidth,keepaspectratio]{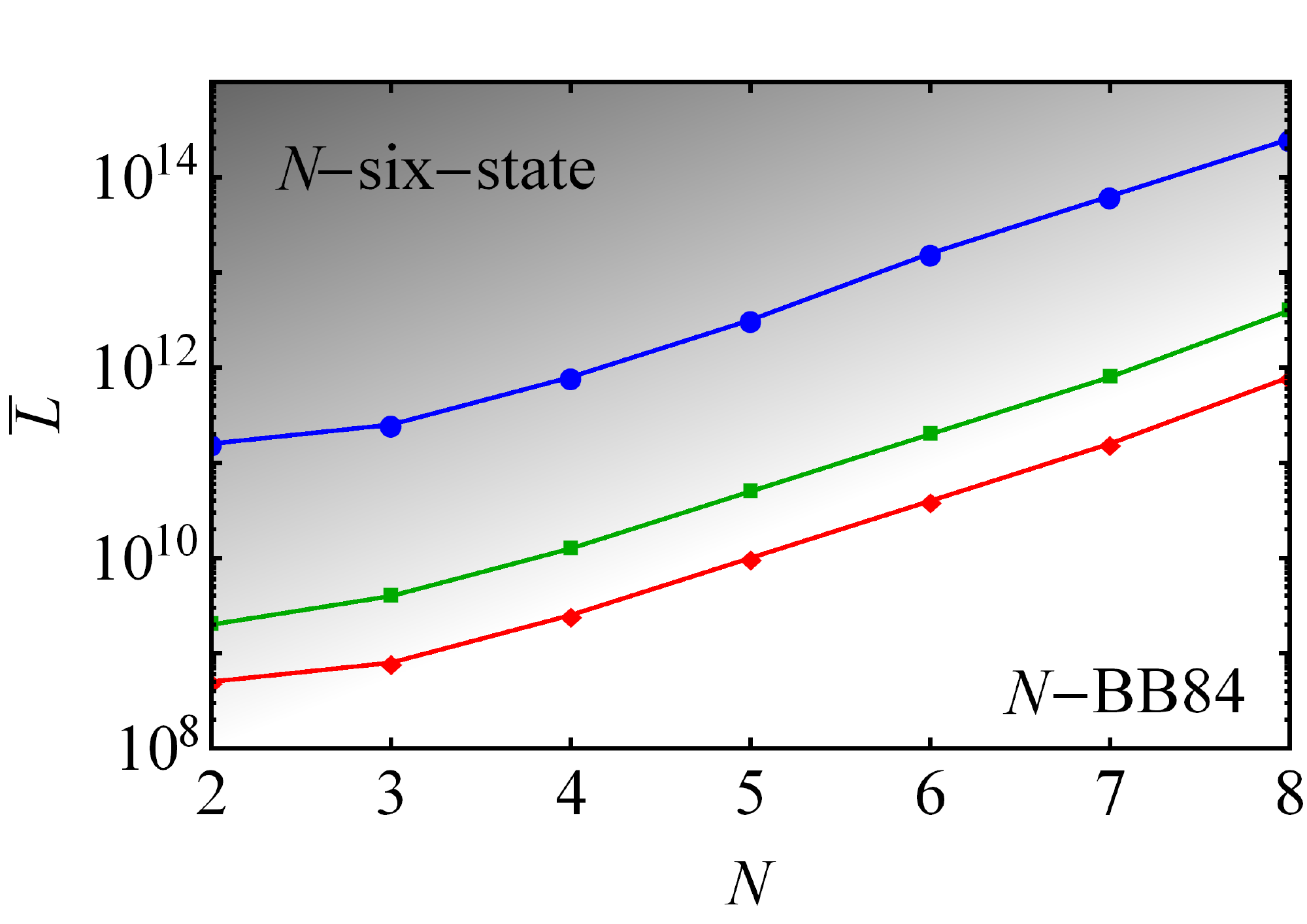}
		\caption{Threshold function $\bar{L}$ for \\ \mbox{$Q^m_{AB}=0.01,0.05,0.1$} (blue circles, green squares, red diamonds) 
			as a function of the number of parties $N$.}
		\label{LbarfixedQ}
	\end{subfigure}%
	\begin{subfigure}[t]{.5\textwidth}
		\includegraphics[width=1\linewidth,keepaspectratio]{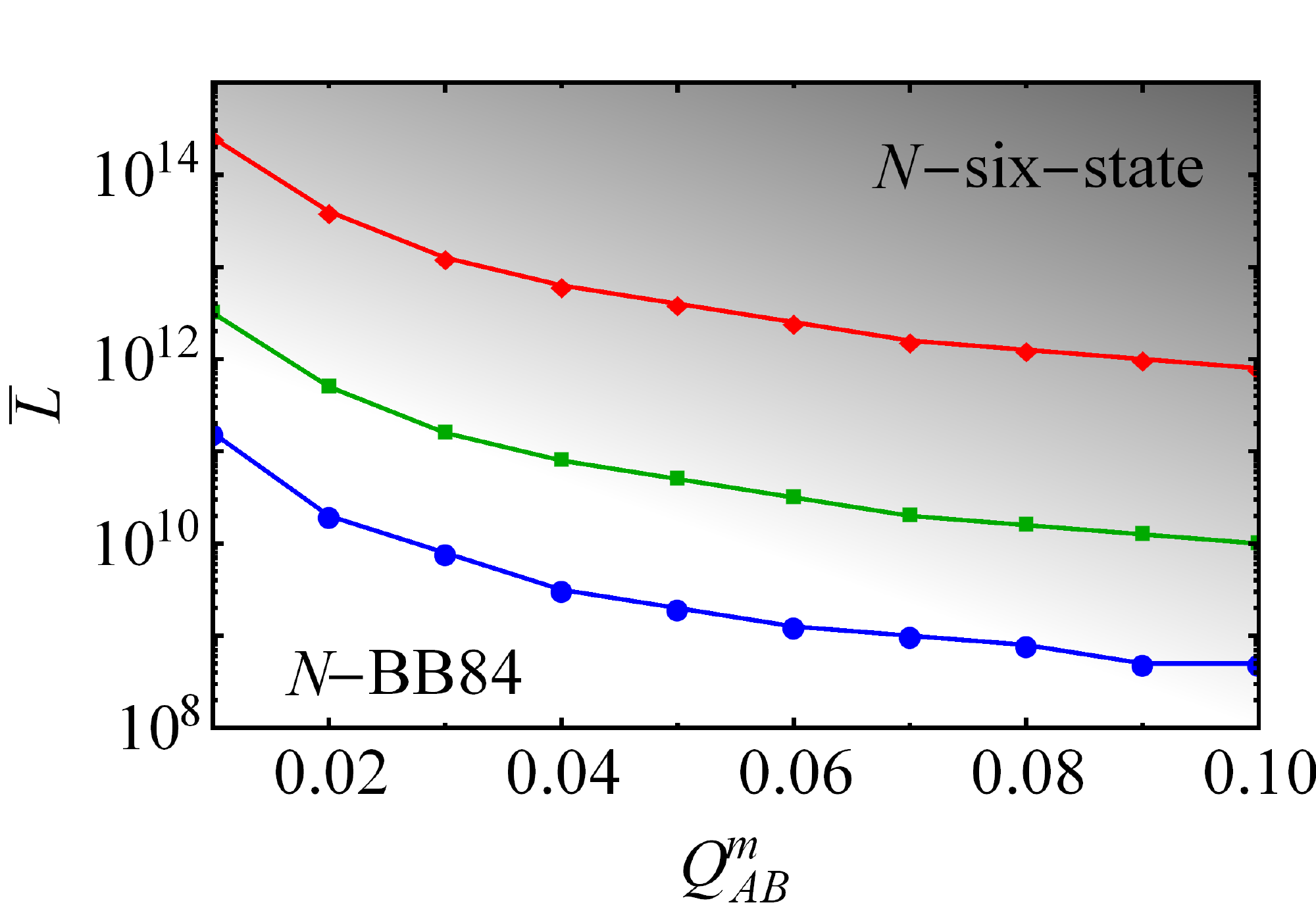}
		\caption{Threshold function $\bar{L}$ for \mbox{$N=2,5,8$} (blue circles, green squares, red diamonds) as a function of $Q^m_{AB}$, 
			proportional to the channel noise.}
		\label{LbarfixedN}
	\end{subfigure}
	\caption{The threshold $\bar{L}$ as a function of one of its variables, while keeping the other one fixed.}
	\label{thresholdfunction}
\end{figure}
In \autoref{finite-rates} we compare the key rates of both NQKD protocols for a finite number of signals $L$ 
transmitted through the quantum channel, with noise discussed in \autoref{errormodel}. The rates are numerically maximized over the parameters:
$p,\bar{\varepsilon},\varepsilon_{\mathrm{PE}},\varepsilon_{\mathrm{EC}},\varepsilon_{\mathrm{PA}}$, with the constraint given by the fixed value of the 
total security parameter: $\varepsilon_{\mathrm{tot}}=5\cdot 10^{-9}$. The fact that we are still able to obtain non-zero rates in the finite-key
scenario means that the correction term ``$-h(p)$'' due to the preshared secret key is not prominent, as a matter of fact the optimal values for $p$ are typically well
below $0.1$.\\
We observe that, although for large $L$ the $N$-six-state still performs better than the $N$-BB84 protocol, there exists a certain number of rounds -identified 
by the threshold function $\bar{L}(Q^m_{AB},N)$- below which the $N$-six-state protocol is outperformed by the $N$-BB84 protocol.
The threshold function $\bar{L}$ is defined as:
\begin{equation}
\fl \bar{L}(Q^m_{AB},N) = \min L \quad s.t. \quad r_{6\mbox{\footnotesize-state}}(L,Q^m_{AB},N) \geq r_{\mathrm{BB84}}(L,Q^m_{AB},N)  \label{Lbar}\,\, .
\end{equation}
From \autoref{fixedQincreasingN} one deduces that the $N$-six-state protocol is much more sensitive than the $N$-BB84 if the number of parties is increased,
displaying the opposite behavior with respect to the asymptotic case (\autoref{asymptotic_plot1}).
This causes the threshold function to increase with $N$ and fixed $Q^m_{AB}$ (\autoref{LbarfixedQ}). \\
On the other hand, the $N$-six-state protocol is more robust than the $N$-BB84 protocol when the quantum channels become noisier (\autoref{fixedNincreasingQ}).
As a result the threshold function decreases for increasing noise and fixed $N$ (\autoref{LbarfixedN}).\\
We point out that the function $\bar{L}$ may not be a physical threshold for the number of rounds above which the $N$-six-state protocol is more efficient 
than the $N$-BB84 protocol, as the achievable key rates depend on quantitatively different estimates.
As a matter of fact, it is known \cite{uncertainty-rel} that the uncertainty relation employed for the $N$-BB84 protocol yields tighter
bounds compared to the PS technique used for the $N$-six-state protocol, especially for low values of $L$. Instead, asymptotically 
the correction terms introduced by the PS technique and the uncertainty relation 
vanish\footnote[1]{Recall that the correction terms due to PS allow to extend the security of the key
against collective attacks to coherent attacks, however in the asymptotic limit these attacks are equivalent \cite{scarani-renner-2008}, thus the PS 
corrections vanish.}, allowing the $N$-six-state to outperform the $N$-BB84 protocol (\autoref{asymptotic_plot1}). Therefore the crossover between the two key rates
at $\bar{L}$ is mainly caused by the different tightness of the min-entropy bounds used in the two protocols.\\
Moreover, the PS corrections become more pronounced for increasing number of parties, thus explaining the rise of the threshold function with $N$.
Indeed, the reduction in the key length scales quadratically with the dimension $d$ of the Hilbert space of a single-signal state shared by all $N$ parties. 
Since we assume that the quantum system held by each party
is a qubit, $d=2^N$, i.e. the reduction in the key length introduced by the PS technique scales exponentially in~$N$.

\subsection{Why different strategies?}
In \autoref{finiteresources} we argued that the $N$-BB84 protocol outperforms the $N$-six-state protocol, at low values of $L$, due to the adoption of tighter bounds
on the min-entropy. One could wonder what would happen if the same strategy were used in obtaining the computable key length for both protocols.
Unfortunately, this is not possible: the two strategies employed (uncertainty relation and PS technique) are suited to the particular protocol to which they are applied 
and they cannot be used in the other protocol.\\
In principle the uncertainty relation may also be used to bound the min-entropy of the $N$-six-state protocol, but then the additional symmetries due to the extended
depolarization procedure would be ignored, such that one ends with the same key length as for the $N$-BB84 protocol.\\
Conversely, one could employ the PS technique in combination with the AEP to bound the min-entropy of the $N$-BB84 protocol. The problem in this case would be the
lack of information provided by any symmetrization procedure performed on the shared signals. Indeed without any further symmetrization, the degrees of freedom of the
shared signals\footnote{Remember that we are considering $N$-qubit states, thus their degrees of freedom are much more than in the bipartite case.}, reduced by the 
PE observations, would still be too many to find a computable bound to the min-entropy (i.e. a bound that only depends on the PE statistics and on the input parameters).

\section{Conclusion and Outlook}  \label{conclusion}
In this paper we presented the first complete finite-key analysis of two $N$-partite QKD (NQKD) protocols, which can be regarded as the multipartite versions
of the BB84 \cite{BB84} and of the six-state \cite{Bruss} protocol. Although both protocols adopt genuinely multipartite entangled states as resources, these states are 
only required for a small number of rounds, while in the majority of the cases product states are distributed.\\
In order to study finite-size effects in NQKD schemes, 
we extended the information theoretic security analysis \cite{RennerThesis} of bipartite QKD protocols to the multipartite case, taking into
account both one-way and two-way error correction protocols.
Then we employed the general results on the security of NQKD to investigate the $N$-six-state protocol \cite{Epping} and the newly-defined $N$-BB84 protocol. 
In particular, we derived analytical formulas for the achievable secret key length of both protocols which only depend on the parameter estimation statistics and on the 
desired level of security. We achieved this by bounding the knowledge of the eavesdropper about the secret key by means of the best-known strategies
adopted in bipartite QKD, namely the uncertainty relation for smooth entropies \cite{uncertainty-rel} and the postselection technique \cite{postsel}.\\
We compared the performance of the two NQKD protocols in the case of finite resources and in the asymptotic limit.
We observed that, although the $N$-six-state protocol reaches higher rates asymptotically, there exists a threshold value for the number of signals below which it is
outperformed by the $N$-BB84 protocol. We argued that this crossover between the rates of the two protocols is caused by the different strategies adopted in obtaining the 
computable key lengths, and we justified the choice of the strategy for each protocol.\medskip\\
In order to carry out a fairer comparison between the $N$-six-state protocol and the \mbox{$N$-BB84} when the number of available resources is low, it would
be desirable to implement tighter bounds for the min-entropy of the former protocol.
In any case, the framework of NQKD $\varepsilon$-security developed in this paper may be used for the finite-key analysis of other multipartite QKD
protocols.\\
This work is based on the assumptions that the measurement devices are ideal and that the parties have access to true randomness. In order to address more realistic
scenarios, one can consider the fact that the measurements in the $Z$ and $X$ bases are not
necessarily projective measurements in diagonal bases, but rather generic positive operator-valued measurements. 
This fact could be easily implemented in our $N$-BB84 protocol, thanks to the properties of the uncertainty relation \cite{TomamichelNature}. A more drastic approach
is represented by device-independent QKD (DIQKD) \cite{VaziraniVidick,FriedmanRennerNature}, where no assumption is made on the devices except for spatial separation.
In this context it is worth mentioning the recent security proof of a multipartite DIQKD protocol \cite{DICKA}.
In that protocol security is guaranteed for every violation of a bipartite Bell inequality (CHSH inequality \cite{CHSH}) between one of the parties and the other $N-1$.
It is not yet known whether security can still be proven for violations of a multipartite Bell inequality (MABK inequality \cite{MABK1,MABK2,MABK3}) 
that do not necessarily imply CHSH violations.

\ack
We thank Renato Renner for having generously provided the proof of the result presented in \ref{RennerThm}.
This project has received funding from the European Union’s Horizon 2020 research and innovation programme under the Marie Skłodowska-Curie grant agreement No 675662
and support from the Federal Ministry of Education and Research (BMBF, projects Q.com-Q and HQS).

	\appendix
	\section{Notation}  \label{notation}
	\begin{itemize}
		\item The binary entropy function is defined as: $h(p)=-p\log_2 p - (1-p) \log_2(1-p)$, for $p\in[0,1]$.
		\item The norm $\norm{\cdot}$ of an operator $O$ is defined as: $\norm{O}=\mathrm{Tr}[\sqrt{O^{\dag}O}]$.
		\item $\mathcal{P}(\mathcal{H})$ is the set of positive-semidefinite operators on the Hilbert space $\mathcal{H}$.
		\item The set of possible secret keys shared by the parties is $\mathcal{S}$.
		\item The set of operators which are $\varepsilon$-close to a given density operator $\rho$ is defined as:
		\begin{equation}
		\mathcal{B}^\varepsilon(\rho) \equiv \left\lbrace \left. \tau \in \mathcal{P}(\mathcal{H}) \, : \, \Tr[\tau]\leq 1 \,,\,\frac{1}{2}\norm{\tau - \rho} 
		\leq \varepsilon \right. \right\rbrace  \label{epsilon-close-trace}
		\end{equation}
		if the distance is computed with respect to the trace distance, or as:
		\begin{equation}
		\mathcal{B}^{\varepsilon,\,\mathrm{P}}(\rho) \equiv
		\{\tau \in \mathcal{P}(\mathcal{H}) \, : \, \Tr[\tau]\leq 1 \,,\, P(\tau,\rho) \leq \varepsilon\}  \label{epsilon-close-purified}
		\end{equation} 
		if the distance is given by the purified distance \cite{purifieddistance}:
		\begin{eqnarray}
		  P(\tau,\rho) \equiv \sqrt{1-{\bar{F}(\tau,\rho)}^2} \nonumber 
		\end{eqnarray}
		where $\bar{F}(\tau,\rho)$ is called generalized fidelity:
		\begin{equation}
			\bar{F}(\tau,\rho) \equiv \Tr|\sqrt{\tau}\sqrt{\rho}| + \sqrt{(1-\Tr \rho)(1-\Tr \tau)}  \,\,. \nonumber
		\end{equation}
		Since the purified distance is an upper bound to the trace distance \cite{purifieddistance}, it holds:
		\begin{equation}
		\mathcal{B}^{\varepsilon,\,\mathrm{P}}(\rho) \subseteq \mathcal{B}^\varepsilon(\rho) \quad. \label{Bpurified-in-B}
		\end{equation}
		\item We say that $\rho_X$ is the operator representation of the probability distribution $P_X$ on the set $\mathcal{X}$ if:
		\begin{equation}
			\rho_X \equiv \sum_{x\in\mathcal{X}} P_X(x) \ket{x}\bra{x}   \label{operator-representation} 
	    \end{equation}
	    for some orthonormal basis $\{\ket{x}\}_x$.
	    \item We define the set of probability distributions which are $\varepsilon$-close to a given probability distribution $P_X$ as those
	    distributions whose operator representation is $\varepsilon$-close to the operator representation of $P_X$, according to 
	    (\ref{epsilon-close-trace}) and (\ref{epsilon-close-purified}).
		\item The Rényi \mbox{zero-entropy} $H_{0} (P_{XY}|Y)$ of the probability distribution $P_{XY}$ over the set $\mathcal{X}\times\mathcal{Y}$ is given by
		\cite{RennerThesis,AEP}:
		\begin{equation}
			H_{0} (P_{XY}|Y) \equiv \log_2 \max_{y\in\mathcal{Y}} |\mathrm{supp}(P_X^y)|   \label{Renyizero}
		\end{equation}
		where $P_X^y$ denotes the function $P_X^y : x\mapsto P_{XY}(x,y)$. This entropy was called ``max-entropy'' in \cite{RennerThesis}.
		\item The $\varepsilon$-smooth Rényi \mbox{zero-entropy} $H_{0}^{\varepsilon} (P_{XY}|Y)$ is defined as \cite{RennerThesis, RennerAndWolf}:
		\begin{equation}
		  	H_{0}^{\varepsilon} (P_{XY}|Y) \equiv \min_{Q_{XY}\in\mathcal{B}^\varepsilon(P_{XY})} H_{0} (Q_{XY}|Y) \,\,.  \label{smoothRenyizero}
		\end{equation}
		If the minimization is performed on $\mathcal{B}^{\varepsilon,\,\mathrm{P}}(P_{XY})$ the corresponding Rényi \mbox{zero-entropy}
		is denoted as: $H_{0}^{\varepsilon,\,\mathrm{P}}(P_{XY}|Y)$.
		\item The Rényi \mbox{zero-entropy} $H_{0} (\rho)$ of the density operator $\rho$ is defined as \cite{RennerThesis}:
		\begin{equation}
			H_{0} (\rho) \equiv \log_2 \mathrm{rank}(\rho)  \label{Renyizeroquantum}
		\end{equation}
		\item The min-entropy of the density operator $\rho_{AB}$ relative to $\sigma_B$ is \cite{RennerThesis,AEP}:
		\begin{equation}
		\fl H_{\mathrm{min}}(\rho_{AB}|\sigma_B) \equiv -\log_2 \min \set{\lambda\in\mathbb{R}:\, \lambda(\mathrm{id}_A \otimes \sigma_B) -\rho_{AB} \geq 0}
		\label{min-entropy-rel}
		\end{equation}
		Note that for $H_{\mathrm{min}}(\rho_{AB}|\sigma_B)$ to exist, a necessary condition is that: \mbox{$\mathrm{supp}(\rho_B)\subseteq
		\mathrm{supp}(\sigma_B)$}. If $\mathcal{H}_B$ is the trivial space $\mathbb{C}$, then the min-entropy reduces to:
		\begin{equation}
			H_{\mathrm{min}}(\rho_{A}) = -\log_2 \lambda_{\mathrm{max}} (\rho_A)  \label{min-entropy-trivial} 
		\end{equation}
		where $\lambda_{\mathrm{max}} (\rho_A)$ is the maximum eigenvalue of $\rho_A$.
	    \item The min-entropy of the probability distribution $P_{XY}$ relative to the distribution $Q_Y$ is \cite{RennerThesis}:
	    \begin{equation}
	    	H_{\mathrm{min}}(P_{XY}|Q_Y) \equiv H_{\mathrm{min}}(\rho_{XY}|\sigma_Y)  \label{min-entropy-probability}
	    \end{equation}
	    where $\rho_{XY}$ and $\sigma_Y$ are the operators representations (\ref{operator-representation}) of $P_{XY}$ and $Q_Y$, respectively.
		\item The min-entropy of $A$ conditioned on $B$ of the density operator $\rho_{AB}$ is \cite{RennerThesis,AEP,TomamichelBook}:
		\begin{equation}
		\fl H_{\mathrm{min}}(\rho_{AB}|B) \equiv -\log_2 \min \{\mathrm{Tr}\sigma_B : \sigma_B\in \mathcal{P}(\mathcal{H}_B) \,,\, 
		(\mathrm{id}_A \otimes \sigma_B) -\rho_{AB} \geq 0 \}  \label{min-entropy}
		\end{equation}
		\item The $\varepsilon$-smooth min-entropy of $A$ conditioned on $B$ of the state $\rho_{AB}$ is \cite{RennerThesis,AEP}:
		\begin{equation}
		H_{\mathrm{min}}^{\varepsilon}(\rho_{AB}|B)\equiv \max_{\tilde{\rho}_{AB}\in\mathcal{B}^{\varepsilon}(\rho_{AB})} H_{\mathrm{min}}(\tilde{\rho}_{AB}|B) \,\,.
		\label{smooth-min-entropy}
		\end{equation}
		If the maximization is performed on $\mathcal{B}^{\varepsilon,\,\mathrm{P}}(\rho_{AB})$ the corresponding min-entropy is denoted as:
		$H_{\mathrm{min}}^{\varepsilon,\,\mathrm{P}}(\rho_{AB}|B)$.
		\item The max-entropy of $A$ conditioned on $B$ of the density operator $\rho_{AB}$ is \cite{AEP}:
		\begin{equation}
		H_{\mathrm{max}}(\rho_{AB}|B)\equiv -H_{\mathrm{min}}(\rho_{AC}|C) \label{max-entropy}
		\end{equation}
		where the min-entropy of the r.h.s. is evaluated for a purification $\rho_{ABC}$ of $\rho_{AB}$.
		\item The $\varepsilon$-smooth max-entropy of $A$ conditioned on $B$ of the density operator $\rho_{AB}$ is \cite{AEP}:
		\begin{equation}
		H_{\mathrm{max}}^\varepsilon(\rho_{AB}|B)\equiv \min_{\tilde{\rho}_{AB}\in\mathcal{B}^{\varepsilon}(\rho_{AB})} H_{\mathrm{max}}(\tilde{\rho}_{AB}|B) \,\,.
		\label{smooth-max-entropy}
		\end{equation}
		If the minimization is performed on $\mathcal{B}^{\varepsilon,\,\mathrm{P}}(\rho_{AB})$ the corresponding max-entropy is denoted as:
		$H_{\mathrm{max}}^{\varepsilon,\,\mathrm{P}}(\rho_{AB}|B)$.
	\end{itemize}
	
	\section{Further NQKD definitions and theorems' proofs}  \label{NQKD_def_and_proofs}
	In this appendix we prove the two results (\autoref{achievable1and2way} and \autoref{leak_upperbound}) presented in \autoref{MultipartiteQKD}.\\
	First we show that correctness and secrecy of a protocol are a sufficient condition for security (\autoref{def-secure}), analogously to the bipartite case
	\cite{RennerThesis,Portman}:
	\begin{defn} \label{def-correctness}
		{\normalfont \cite{DICKA},\cite{TomamichelNature}}.
		Let $\rho_{A\mathbf{B}E}$ be a density operator. Any NQKD protocol, which is $\varepsilon_{\mathrm{rob}}$-robust
		on $\Tr_E[\rho_{A\mathbf{B}E}]$, is said to be {\normalfont $\varepsilon'$-correct} on $\rho_{A\mathbf{B}E}$ if:
		\begin{equation}
		(1-\varepsilon_{\mathrm{rob}}) \mathrm{Pr}\left[\exists i\in\{1,\dots,N-1\} : s_A\neq s_{B_i}\right]\leq \varepsilon' \nonumber
		\end{equation}
		where $(s_A,\mathbf{s_B})$ are the secret keys generated by the NQKD protocol and the probability is conditioned on the fact that the protocol did not abort.
	\end{defn}
	Note that the definition of robustness of an NQKD protocol is given in \autoref{def-robust}.
	\begin{defn}  \label{def-secrecy}
		{\normalfont \cite{DICKA},\cite{TomamichelNature}}.
		Let $\rho_{A\mathbf{B}E}$ be a density operator. Any NQKD protocol, which is $\varepsilon_{\mathrm{rob}}$-robust
		on $\Tr_E[\rho_{A\mathbf{B}E}]$, is said to be {\normalfont $\varepsilon''$-secret} on $\rho_{A\mathbf{B}E}$ if:
		\begin{equation}
		(1-\varepsilon_{\mathrm{rob}}) \frac{1}{2}\norm{\rho_{S_A E'}- \rho_U \otimes \rho_{E'}}\leq \varepsilon''   \nonumber
		\end{equation}
		where $\rho_U$ is the uniform state on $A$'s key space.
	\end{defn}
	The following lemma holds:
	\begin{lemma}
		Given an NQKD protocol which is $\varepsilon'$-correct and $\varepsilon''$-secret, then it is also $(\varepsilon'+\varepsilon'')$-secure.
		\label{correctness+secrecy}
	\end{lemma}
	
	\textit{Proof}. From the correctness hypothesis we have:
		\begin{eqnarray}
		&\fl\mathrm{Pr}\left[\exists i\in\{1,\dots,N-1\}: s_A\neq s_{B_i}\right] = 1-\mathrm{Pr}\left[\nexists i\in\{1,\dots,N-1\}: s_A\neq s_{B_i}\right] = \nonumber \\
		&\fl= 1- \sum_{s\in\mathcal{S}} P_{S_A \mathbf{S_B}}(s,\dots,s) = 1- \sum_{s_A,\mathbf{s_B}} P_{S_A \mathbf{S_B}}(s_A,\mathbf{s_B}) \delta_{s_A \mathbf{s_B}} 
		\nonumber
		\end{eqnarray}
		where $\delta_{s_A \mathbf{s_B}} \equiv \Pi_{i=1}^{N-1} \delta_{s_A s_{B_i}}$. Therefore, $\varepsilon'$-correctness yields:
		\begin{equation}
		\sum_{s_A,\mathbf{s_B}} P_{S_A \mathbf{S_B}}(s_A,\mathbf{s_B}) (1-\delta_{s_A \mathbf{s_B}}) \leq \frac{\varepsilon'}{1-\varepsilon_{\mathrm{rob}}}  
		\,\,. \label{correctnesshyp}
		\end{equation}
		From the secrecy hypothesis we have:
		\begin{eqnarray}
		&\fl\frac{1}{2}\norm{\rho_{S_A E'}-\rho_U \otimes \rho_{E'}} =  \frac{1}{2}\norm{\sum_{s_A,\mathbf{s_B}}P_{S_A \mathbf{S_B}}(s_A,\mathbf{s_B}) \ket{s_A}\bra{s_A}
			\otimes\rho_{E'}^{s_A,\mathbf{s_B}} - \sum_{s_A}\frac{1}{|\mathcal{S}|}\ket{s_A}\bra{s_A} \otimes \rho_{E'} }   \nonumber  \\
		&\fl= \frac{1}{2}\norm{\sum_{s_A} \ket{s_A}\bra{s_A} \otimes \left(\sum_{\mathbf{s_B}} P_{S_A \mathbf{S_B}}(s_A,\mathbf{s_B}) \rho_{E'}^{s_A,\mathbf{s_B}} - 
			\frac{1}{|\mathcal{S}|} \rho_{E'}\right)}  \nonumber \\
		&\fl= \frac{1}{2}\sum_{s_A} \norm{\sum_{\mathbf{s_B}} P_{S_A \mathbf{S_B}}(s_A,\mathbf{s_B}) \rho_{E'}^{s_A,\mathbf{s_B}} - \frac{1}{|\mathcal{S}|} \rho_{E'}}
		\leq \frac{\varepsilon''}{1-\varepsilon_{\mathrm{rob}}} \,\,. \label{secrecyhyp}
		\end{eqnarray} \bigskip
		
		\noindent Having obtained inequalities (\ref{correctnesshyp}) and (\ref{secrecyhyp}), we are ready to prove the thesis:
		\begin{eqnarray}
		&\fl\frac{1}{2}\norm{\rho_{S_A \mathbf{S_B}E'} - \rho_{\mathbf{U}} \otimes \rho_{E'} } =   \nonumber  \\
		&\fl=\frac{1}{2} \Bigg\lVert \sum_{s_A,\mathbf{s_B}}P_{S_A \mathbf{S_B}}(s_A,\mathbf{s_B}) \ket{s_A}\bra{s_A} \otimes \ket{\mathbf{s_B}}\bra{\mathbf{s_B}}
		\otimes \rho_{E'}^{s_A,\mathbf{s_B}} \nonumber \\ 
		&\fl\hspace{60pt}- \sum_{s_A,\mathbf{s_B}} \frac{1}{|\mathcal{S}|} \delta_{s_A \mathbf{s_B}} \ket{s_A}\bra{s_A}\otimes \ket{\mathbf{s_B}}\bra{\mathbf{s_B}} 
		\otimes \rho_{E'} \Bigg\rVert    \nonumber \\ 
		&\fl= \frac{1}{2} \sum_{s_A,\mathbf{s_B}} \norm{P_{S_A \mathbf{S_B}}(s_A,\mathbf{s_B})\rho_{E'}^{s_A,\mathbf{s_B}} - \frac{\delta_{s_A \mathbf{s_B}}}{|\mathcal{S}|}
			\rho_{E'} } \nonumber  \\
		&\fl= \frac{1}{2} \left[\sum_{s_A,\mathbf{s_B}} (1-\delta_{s_A \mathbf{s_B}}) \norm{P_{S_A \mathbf{S_B}}(s_A,\mathbf{s_B})\rho_{E'}^{s_A,\mathbf{s_B}} -
			\frac{\delta_{s_A \mathbf{s_B}}}{|\mathcal{S}|} \rho_{E'} } \right.  \nonumber  \\
		&\fl\hspace{41.2pt}+ \left. \sum_{s_A,\mathbf{s_B}} \delta_{s_A \mathbf{s_B}} \norm{P_{S_A \mathbf{S_B}}(s_A,\mathbf{s_B})\rho_{E'}^{s_A,\mathbf{s_B}} -
			\frac{\delta_{s_A \mathbf{s_B}}}{|\mathcal{S}|} \rho_{E'} } \right]  \nonumber \\
		&\fl= \frac{1}{2} \left[\sum_{s_A,\mathbf{s_B}} (1-\delta_{s_A \mathbf{s_B}}) \norm{P_{S_A \mathbf{S_B}}(s_A,\mathbf{s_B})\rho_{E'}^{s_A,\mathbf{s_B}}}\right. + 
		 \left. \sum_{s_A}\norm{P_{S_A \mathbf{S_B}}(s_A,\dots,s_A)\rho_{E'}^{s_A,\dots,s_A} - \frac{1}{|\mathcal{S}|} \rho_{E'} } \right]  \nonumber \\
		&\fl\stackrel{(1)}{\leq} \frac{\varepsilon'}{2(1-\varepsilon_{\mathrm{rob}})} + \frac{1}{2}\sum_{s_A}\norm{P_{S_A \mathbf{S_B}}(s_A,\dots,s_A)\rho_{E'}^{s_A,\dots,s_A} -
			\frac{1}{|\mathcal{S}|} \rho_{E'} }  \nonumber   \\
		&\fl\stackrel{(2)}{\leq} \frac{\varepsilon'}{2(1-\varepsilon_{\mathrm{rob}})} + \frac{1}{2}\sum_{s_A}\norm{P_{S_A \mathbf{S_B}}(s_A,\dots,s_A)\rho_{E'}^{s_A,\dots,s_A} -
			\sum_{\mathbf{s_B}} P_{S_A \mathbf{S_B}}(s_A,\mathbf{s_B}) \rho_{E'}^{s_A,\mathbf{s_B}}}   \nonumber  \\
		&\fl\hspace{27pt}  + \frac{1}{2}\sum_{s_A}\norm{\sum_{\mathbf{s_B}} P_{S_A \mathbf{S_B}}(s_A,\mathbf{s_B}) \rho_{E'}^{s_A,\mathbf{s_B}} -
			\frac{1}{|\mathcal{S}|} \rho_{E'} }   \nonumber  \\
		&\fl\stackrel{(3)}{\leq} \frac{\varepsilon'}{2(1-\varepsilon_{\mathrm{rob}})} + \frac{\varepsilon''}{1-\varepsilon_{\mathrm{rob}}} +
		\frac{1}{2}\sum_{s_A}\norm{\sum_{\mathbf{s_B}} P_{S_A \mathbf{S_B}}(s_A,\mathbf{s_B})
			\rho_{E'}^{s_A,\mathbf{s_B}} (1- \delta_{s_A \mathbf{s_B}})}   \nonumber  \\
		&\fl\stackrel{(4)}{\leq} \frac{\varepsilon'}{2(1-\varepsilon_{\mathrm{rob}})} + \frac{\varepsilon''}{1-\varepsilon_{\mathrm{rob}}} +
		\frac{1}{2}\sum_{s_A,\mathbf{s_B}}\norm{ P_{S_A \mathbf{S_B}}(s_A,\mathbf{s_B}) 	\rho_{E'}^{s_A,\mathbf{s_B}} (1- \delta_{s_A \mathbf{s_B}})}   \nonumber \\
		&\fl\stackrel{(5)}{\leq} \frac{\varepsilon'}{1-\varepsilon_{\mathrm{rob}}} + \frac{\varepsilon''}{1-\varepsilon_{\mathrm{rob}}}
		\end{eqnarray}
		which concludes the proof according to the security definition in \autoref{def-secure}. 
		Note that we made use of the following properties: $(1)$ the fact that the operator $\rho_{E'}^{s_A,\mathbf{s_B}}$ is normalized and 
		(\ref{correctnesshyp}); $(2)$ triangle inequality; $(3)$ (\ref{secrecyhyp}); $(4)$ triangle inequality; \mbox{$(5)$ $\rho_{E'}^{s_A,\mathbf{s_B}}$} is normalized and 
		(\ref{correctnesshyp}). \hfill\opensquare\bigskip\\
		We now prove the result on the achievable key length of a general NQKD protocol:\medskip\\
		\textit{Proof of \autoref{achievable1and2way}}.  In the post-processing protocol $\mathsf{PP}_{\{\mathsf{EC}_i\},\mathcal{F}}$, the sub-protocol which transforms
		partially correlated key pairs into fully correlated ones is defined by the set $\{\mathsf{EC}_i\}_{i=1}^{N-1}$. 
		Because $\{\mathsf{EC}_i\}_{i=1}^{N-1}$ is $\varepsilon_{\mathrm{EC}}$-secure (in the sense of \autoref{def-IRsecurity}) on the
		classical probability distribution defined by $\rho_{X\mathbf{K}}$, according to \autoref{def-correctness} the whole NQKD protocol is 
		$\varepsilon_{\mathrm{EC}}$-correct on $\rho_{A\mathbf{B}E}$. Thus by \autoref{correctness+secrecy} we only need to show that the NQKD protocol
		is $(2\bar{\varepsilon}+\varepsilon_{\mathrm{PA}})$-secret in order to complete the proof, i.e. :
		\begin{equation}
		\frac{1}{2}\norm{\rho_{S_A E'}- \rho_U \otimes \rho_{E'}}\leq \frac{2\bar{\varepsilon}+\varepsilon_{\mathrm{PA}}}{1-\varepsilon_{\mathrm{rob}}}   \,\,. \label{secrecytoprove}
		\end{equation}
		We stress the fact that in Eve's subsystem $E'$ we included not only Eve's quantum degree of freedom $\mathcal{H}_E$, but also her knowledge about the classical
		communication $\mathcal{H}_{\mathbf{C}}$ occurring during error correction (defined by $\{\mathsf{EC}_i\}$) and the classical communication taking place in privacy
		amplification $\mathcal{H}_F$ (defined by the set $\mathcal{F}$).\\ 
		In order to prove (\ref{secrecytoprove}), we start from the result in \cite[Corollary 5.6.1]{RennerThesis} stated in a slightly weaker form:
		\begin{equation}
		\norm{\rho_{S_A E'}- \rho_U \otimes \rho_{E'}} \leq \frac{4\bar{\varepsilon}\,'}{1-\varepsilon_{\mathrm{rob}}} + 
		2^{-\frac{1}{2}\left(H_{\mathrm{min}}^{\bar{\varepsilon}\,'}(\rho_{X \mathbf{C}E}|\mathbf{C}E) -\ell \right)} \label{corollary561}
		\end{equation}
		valid $\forall \,\, \bar{\varepsilon}\,'$, where $\ell$ is the number of key bits after privacy amplification. 
		The inequality (\ref{corollary561}) leads to a sufficient condition for (\ref{secrecytoprove}) to be true, namely:
		\begin{equation}
		H_{\mathrm{min}}^{\bar{\varepsilon}\,'}(\rho_{X \mathbf{C}E}|\mathbf{C}E) -\ell \geq 2\log_2 
		\frac{1-\varepsilon_{\mathrm{rob}}}{2(2\bar{\varepsilon}+\varepsilon_{\mathrm{PA}} -2\bar{\varepsilon}\,' )}	\label{ineqtoprove}
		\end{equation}
		therefore we will now focus on proving (\ref{ineqtoprove}), having fixed: $\bar{\varepsilon}\,' = \bar{\varepsilon}$. \\
		We first prove the result without assuming that the classical communication \textbf{C} is one-way, i.e. it may also depend on \textbf{B}'s raw keys. Then
		we show how to achieve a slightly stronger result by assuming one-way classical communication.\medskip\\
		\textbf{TWO-WAY EC:} Since the purified distance is an upper bound to the trace distance, an $\varepsilon$-environment defined with
		the latter is larger (\ref{Bpurified-in-B}). Thus:
		\begin{eqnarray}
		H_{\mathrm{min}}^{\bar{\varepsilon}}(\rho_{X \mathbf{C}E}|\mathbf{C}E) &\geq 
		H_{\mathrm{min}}^{\bar{\varepsilon},\, \mathrm{P}}(\rho_{X \mathbf{C}E}|\mathbf{C}E) \,\,.\label{eq1}
		\end{eqnarray}
		The result stated in \ref{RennerThm} yields:
		\begin{equation}
		H_{\mathrm{min}}^{\bar{\varepsilon},\, \mathrm{P}}(\rho_{X \mathbf{C}E}|\mathbf{C}E) \geq H_{\mathrm{min}}^{\bar{\varepsilon},\,\mathrm{P}}(\rho_{X E}|E)  
		- \left(H_{0}(\rho_{\mathbf{C}}) - H_{\mathrm{min}}(\rho_{X\mathbf{K}\mathbf{C}}|\rho_{X\mathbf{K}})\right)
		\,\,. \label{eq2}
		\end{equation}
		Now let us concentrate on the last two terms in (\ref{eq2}):
		\begin{enumerate}
			\item By definition (\ref{Renyizeroquantum}): $H_{0}(\rho_{\mathbf{C}})=\log_2 \mathrm{rank} (\rho_{\mathbf{C}})$, with:
			\begin{equation}
			\rho_{\mathbf{C}}= \sum_{c_1,\dots,c_{N-1}} P_{\mathbf{C}}(c_1,\dots,c_{N-1})\bigotimes_{i=1}^{N-1} \ket{c_i}\bra{c_i} \,\, , \nonumber
			\end{equation}
			therefore $\mathrm{rank}(\rho_{\mathbf{C}})=|\mathcal{C}_{1,\dots,N-1}|$ according to (\ref{setofcommunications}).
			\item By definition (\ref{min-entropy-rel}):
			$H_{\mathrm{min}}(\rho_{X\mathbf{K}\mathbf{C}}|\rho_{X\mathbf{K}})=-\log_2 \min\lambda$, where $\lambda$ is a real parameter satisfying:
			\begin{eqnarray}
			&\lambda (\rho_{X\mathbf{K}}\otimes \mathrm{id}_{\mathbf{C}}) - \rho_{X\mathbf{KC}} \geq 0  \nonumber \\
			\iff \,\, &\lambda \geq P_{\mathbf{C}|X=x,\mathbf{K}=\mathbf{k}}(c_1,\dots,c_{N-1}|x,\mathbf{k}) \quad \forall x,\mathbf{k},c_1,\dots,c_{N-1} \,\,. \nonumber
			\end{eqnarray}
			Therefore
			\begin{equation}
			\min \lambda = \max_{\mathbf{c},\mathbf{k},x} P_{\mathbf{C}|X=x,\mathbf{K}=\mathbf{k}}(c_1,\dots,c_{N-1}|x,\mathbf{k})\,\, ,  \nonumber
			\end{equation}
			which yields:
			\begin{eqnarray}
			H_{\mathrm{min}}(\rho_{X\mathbf{K}\mathbf{C}}|\rho_{X\mathbf{K}}) &= \min_{x,\mathbf{k}}\left[-\log_2 
			\max_{\mathbf{c}} P_{\mathbf{C}|X=x,\mathbf{K}=\mathbf{k}}(c_1,\dots,c_{N-1}|x,\mathbf{k})\right]  =  \nonumber \\
			&=  \min_{x,\mathbf{k}} H_{\mathrm{min}}(P_{\mathbf{C}|X=x,\mathbf{K}=\mathbf{k}})  \nonumber
			\end{eqnarray}
			where in the last inequality we used the definition of min-entropy for probability distributions (\ref{min-entropy-probability}).
		\end{enumerate}
		Substituting now in (\ref{eq2}), recalling \autoref{def-leak} and using (\ref{eq1}) yields:
		\begin{equation}
		H_{\mathrm{min}}^{\bar{\varepsilon}}(\rho_{X \mathbf{C}E}|\mathbf{C}E) \geq
		H_{\mathrm{min}}^{\bar{\varepsilon},\,\mathrm{P}}(\rho_{X E}|E) - 
		\mathrm{leak}_{\{\mathsf{EC}_i\}}^{\mathrm{NQKD}} \,\,. \nonumber
		\end{equation}
		By using the assumption (\ref{upperbound-2wayIR}) in the last inequality concludes the proof:
		\begin{eqnarray}
		H_{\mathrm{min}}^{\bar{\varepsilon}}(\rho_{X \mathbf{C}E}|\mathbf{C}E)
		&\geq  H_{\mathrm{min}}^{\bar{\varepsilon},\,\mathrm{P}}(\rho_{X E}|E) - 
		\mathrm{leak}_{\{\mathsf{EC}_i\}}^{\mathrm{NQKD}}  \nonumber	\\
		&\geq \ell + 2\log_2 \frac{1-\varepsilon_{\mathrm{rob}}}{2\,\varepsilon_{\mathrm{PA}}} 
		\end{eqnarray}
		since we have just obtained (\ref{ineqtoprove}) with fixed $\bar{\varepsilon}\,' = \bar{\varepsilon}$.\medskip\\
		\textbf{ONE-WAY EC:} For the chain rule \cite[Eq. 3.21]{RennerThesis} we have:
		\begin{equation}
		H_{\mathrm{min}}^{\bar{\varepsilon}}(\rho_{X \mathbf{C}E}|\mathbf{C}E) \geq H_{\mathrm{min}}^{\bar{\varepsilon}}(\rho_{X \mathbf{C}E}|E)-
		H_{0}(\rho_{\mathbf{C}}) \,\,, \label{Eq1}
		\end{equation}
		where the quantum state is, under the assumption of one-way EC protocols:
		\begin{equation}
		\hat{\rho}_{X \mathbf{C}E} = \sum_x \ket{x}\bra{x}\otimes \hat{\rho}_{\mathbf{C}}^x \otimes \rho_E^x  \label{quant-state-oneway}
		\end{equation}
		where the hat $\hat{\cdot}$ indicates normalized density operators and:
		\begin{equation}
		\rho_E^x \equiv \sum_{\mathbf{k}} P_{X\mathbf{K}} (x,\mathbf{k}) \hat{\rho}_E^{x,\mathbf{k}} \,\,.  \nonumber	
		\end{equation}
		Since in (\ref{quant-state-oneway}) the state conditioned on the classical subsystem $\mathcal{H}_X$ is a product state, by \cite[Eq. 3.22]{RennerThesis} we
		conclude that:
		\begin{equation}
		H_{\mathrm{min}}^{\bar{\varepsilon}}(\rho_{X \mathbf{C}E}|E) \geq H_{\mathrm{min}}^{\bar{\varepsilon}}(\rho_{XE}|E) + H_{\mathrm{min}}(\rho_{X \mathbf{C}}|\rho_X)
		\label{Eq2} \,\,.
		\end{equation}
		Substituting (\ref{Eq2}) in (\ref{Eq1}) yields:
		\begin{equation}
		H_{\mathrm{min}}^{\bar{\varepsilon}}(\rho_{X \mathbf{C}E}|\mathbf{C}E) \geq H_{\mathrm{min}}^{\bar{\varepsilon}}(\rho_{XE}|E) - 
		\left(H_{0}(\rho_{\mathbf{C}}) - H_{\mathrm{min}}(\rho_{X \mathbf{C}}|\rho_X) \right)  \label{Eq3}  \,\,,
		\end{equation}
		which is equivalent to what was obtained in the two-way scenario (\ref{eq2}) except for the $\varepsilon$-environment of the min-entropy, here defined via the 
		trace distance. Analogous steps to those employed in the first part lead to the claim valid for one-way EC.\hfill\opensquare\bigskip\\
		Finally, we show how to obtain an upper bound on the leakage of an optimal EC protocol.\medskip\\
		\textit{Proof of \autoref{leak_upperbound}}. Let $\mathcal{X}$ be the set of possible raw keys held by $A$, while $\mathcal{K}$ is the set of possible raw keys
		held by $\mathbf{B}$. Let us consider the following $N$-partite one-way error correction protocol $\mathrm{EC}_{\hat{\mathcal{X}},\mathcal{F}}$
		(generalization of the bipartite version in \cite[Lemma 6.3.3]{RennerThesis}):
		\begin{mdframed}
			Parameters:
			\begin{itemize}
				\item $\hat{\mathcal{X}}$: family of sets $\hat{\mathcal{X}}_{k_i}^i \subseteq \mathcal{X}$ parametrized by the index $i$ which identifies $B_i$ and by $k_i\in
				\mathcal{K}$.
				\item $\mathcal{F}$: family of hash functions from $\mathcal{X}$ to $\mathcal{Z}$.  
			\end{itemize}
			Protocol:
			\begin{enumerate}
				\item $A$ receives as input the raw key $x\in\mathcal{X}$, while $B_i$ receives the raw key $k_i\in\mathcal{K}$.
				\item  $A$ chooses uniformly at random $f\in_R \mathcal{F}$ and defines $z\equiv f(x)$. Then, $A$ sends the classical message $(f,z)$ to $\mathbf{B}$.
				\item $B_i$ selects the set $\hat{\mathcal{X}}_{k_i}^i$ corresponding to the key $k_i$ he is holding, and defines: $\hat{\mathcal{D}}_i\equiv \{\hat{x_i} \in
				\hat{\mathcal{X}}_{k_i}^i : f(\hat{x_i}) = z\}$.
				\item If $\hat{\mathcal{D}}_i \neq \emptyset$ then $B_i$'s guess of $A$'s key is $\hat{x_i}\in_R \hat{\mathcal{D}}_i$, otherwise the protocol aborts.
			\end{enumerate}
		\end{mdframed}
		The proof consists of two parts. The first part extends the result stated in \cite[Lemma 6.3.3]{RennerThesis} to the multipartite scenario, while the second part
		generalizes \cite[Lemma 6.3.4]{RennerThesis}. \\
		\textbf{PART 1:} We first show that the above-defined $\mathrm{EC}_{\hat{\mathcal{X}},\mathcal{F}}$, for an appropriate choice of the parameters
		$\hat{\mathcal{X}}$ and $\mathcal{F}$, is $0$-robust on $P_{X\mathbf{K}}$, $\varepsilon_{\mathrm{EC}}$-fully secure (see \autoref{def-IRsecurity}), and 
		has leakage:
		\begin{equation}
		\mathrm{leak}_{\mathrm{EC}_{\hat{\mathcal{X}},\mathcal{F}}}^{\mathrm{NQKD}}
		\leq \max_i H_{0}(P_{XK_i}|K_i) + \log_2 \left(2/\varepsilon_{\mathrm{EC}}\right) + \log_2 (N-1) \,\,.  \label{intermediate-result}
		\end{equation}
		Let $z_{\mathrm{EC}}\equiv \lceil \max_i H_{0}(P_{XK_i}|K_i) + \log_2 (N-1) + \log_2 (1/\varepsilon_{\mathrm{EC}})\rceil$ and let $\mathcal{F}$ 
		be a two-universal family of hash functions
		from $\mathcal{X}$ to $\mathcal{Z}=\{0,1\}^{z_{\mathrm{EC}}}$. Moreover, let $\hat{\mathcal{X}} = \{\hat{\mathcal{X}}_{k_i}^i\}$ be the family
		of sets defined by $\hat{\mathcal{X}}_{k_i}^i \equiv \mathrm{supp}(P_X^{i,k_i})$, where $\mathrm{supp}(P_X^{i,k_i})$ denotes the support of the function:
		$P_X^{i,k_i} : x\mapsto P_{XK_i}(x,k_i)$.
		From the choice of $\mathcal{F}$ we know that: $\mathrm{Pr}_f [f(x')=f(x)]_{x' \neq x} \leq 2^{-{z_{\mathrm{EC}}}}$ for $f\in_R \mathcal{F}$ 
		and fixed elements $x,x' \in\mathcal{X}$. Note that the two parameters $\hat{\mathcal{X}},\mathcal{F}$ defining the EC protocol 
		are completely fixed by the marginals distributions $P_{XK_i}$ of the given probability distribution $P_{X \mathbf{K}}$.\\
		For any given set of raw keys $(x,k_1,\dots,k_{N-1})$ (not necessarily generated by $P_{X\mathbf{K}}$), one can bound the probability that the protocol 
		$\mathrm{EC}_{\hat{\mathcal{X}},\mathcal{F}}$ does not abort and outputs a wrong guess for at least one Bob, as:
		\begin{eqnarray}\fl
		\mathrm{Pr}_{f,\mathbf{\hat{x}}}\left[\hat{\mathcal{D}}_i \neq \emptyset \,\forall\, i\, \wedge\, \exists i :\hat{x}_i \neq x\right] &\leq 
		\mathrm{Pr}_{f,\mathbf{\hat{x}}}\left[\exists i :\hat{x}_i \neq x\right] \nonumber \\
		&\leq \mathrm{Pr}_{f}\left[\exists \hat{x}\in\cup_{i=1}^{N-1}\hat{\mathcal{D}}_i : \hat{x} \neq x\right] \nonumber \\
		&= \mathrm{Pr}_{f}\left[\exists \hat{x}\in\cup_{i=1}^{N-1}\hat{\mathcal{X}}_{k_i}^i :\hat{x} \neq x \wedge f(\hat{x})=f(x)\right] \nonumber \\
		&\leq \sum_{\hat{x}\in\cup_{i=1}^{N-1}\hat{\mathcal{X}}_{k_i}^i,\hat{x}\neq x} \mathrm{Pr}_f \left[f(\hat{x})\neq f(x) \right] \nonumber \\
		&\leq \sum_{\hat{x}\in\cup_{i=1}^{N-1}\hat{\mathcal{X}}_{k_i}^i,\hat{x}\neq x} 2^{-{z_{\mathrm{EC}}}}  \label{eqn1} 
		\end{eqnarray}
		where the third inequality is due to the union bound and the fourth to the chosen set $\mathcal{F}$. Finally, we can bound (\ref{eqn1}) by:
		\begin{eqnarray}\fl
		\mathrm{Pr}_{f,\mathbf{\hat{x}}}\left[\hat{\mathcal{D}}_i \neq \emptyset \,\forall\, i\, \wedge\, \exists i :\hat{x}_i \neq x\right]
		&\leq \left\vert\cup_{i=1}^{N-1}\hat{\mathcal{X}}_{k_i}^i\right\vert 2^{-{z_{\mathrm{EC}}}} \nonumber \\
		&\leq (N-1) \max_i \max_{k_i} \left\vert\mathrm{supp}(P_X^{i,k_i}) \right\vert 2^{-{z_{\mathrm{EC}}}}\nonumber \\
		&= 2^{\log_2 (N-1)} 2^{\max_i H_{0}(P_{XK_i}|K_i)} 2^{-{z_{\mathrm{EC}}}} \nonumber \\
		&\leq \varepsilon_{\mathrm{EC}} \nonumber  
		\end{eqnarray}
		which proves that $\mathrm{EC}_{\hat{\mathcal{X}},\mathcal{F}}$ is $\varepsilon_{\mathrm{EC}}$-fully secure according to \autoref{def-IRsecurity}. 
		Note that we used (\ref{Renyizero}) for the equality and the definition of ${z_{\mathrm{EC}}}$ in the last inequality. \\
		If the set of keys $(x,k_1,\dots,k_{N-1})$ is now generated by the distribution $P_{X\mathbf{K}}$, then $x\in \hat{\mathcal{X}}_{k_i}^i \,\,\forall i$ since 
		$P_{XK_i}(x,k_i)\neq 0 \,\,\forall i$ (otherwise the pair $(x,k_i)$ could not have been generated). Therefore, being $f(x)=z$ true by definition, the sets
		$\hat{\mathcal{D}}_i$ are never empty, thus the EC protocol never aborts, i.e. it is 0-robust (\autoref{def-robust}) on $P_{X\mathbf{K}}$. \\
		Let us now consider the leakage of the protocol $\mathrm{EC}_{\hat{\mathcal{X}},\mathcal{F}}$. 
	    Since it is a one-way EC protocol where the information sent to one Bob is then copied and then sent to all the other Bobs, the leakage reads (\autoref{def-leak}):
		\begin{equation}
		\mathrm{leak}_{\mathrm{EC}_{\hat{\mathcal{X}},\mathcal{F}}}^{\mathrm{NQKD}}= 
		\log_2 |\mathcal{F}\times \mathcal{Z}| - \min_x H_{\mathrm{min}}(P_{C|X=x}) \,\,.  \label{eqn2} \\
		\end{equation}
		For this EC protocol, after having fixed $A$'s key $x$, the classical communication $(f,z)$ is simply depending on the random choice of $f$, therefore: 
		$P_{C|X=x}=1/|\mathcal{F}|$. Substituting in (\ref{eqn2}) yields:
		\begin{eqnarray}
		\mathrm{leak}_{\mathrm{EC}_{\hat{\mathcal{X}},\mathcal{F}}}^{\mathrm{NQKD}} &= \log_2 |\mathcal{F}\times \mathcal{Z}| - \log_2 |\mathcal{F}| \nonumber \\
		&\leq \log_2 |\mathcal{Z}| ={z_{\mathrm{EC}}} \nonumber \\
		&=  \lceil \max_i H_{0}(P_{XK_i}|K_i) + \log_2 (N-1) + \log_2 (1/\varepsilon)\rceil \nonumber \\
		&\leq \log_2 2 + \max_i H_{0}(P_{XK_i}|K_i) + \log_2 (N-1) + \log_2 (1/\varepsilon) \nonumber \\
		&= \max_i H_{0}(P_{XK_i}|K_i) + \log_2 \left(2/\varepsilon\right) + \log_2 (N-1) \,\,, \nonumber
		\end{eqnarray}
		which concludes the first part of the proof (\ref{intermediate-result}).\medskip\\
		\textbf{PART 2:} Now we employ the result (\ref{intermediate-result}) for another protocol
		$\mathrm{EC}_{\hat{\mathcal{X}},\mathcal{F}}$ where the parameters $\hat{\mathcal{X}},\mathcal{F}$ are defined by a new set of distributions 
		$\set{\bar{P}_{XK_i}}_{i=1}^{N-1}$ linked to the marginals of $P_{X\mathbf{K}}$. Such an EC protocol will be the one that satisfies the claim 
		(\ref{leak-upperbound}).
		The distributions $\set{\bar{P}_{XK_i}}_{i=1}^{N-1}$ are obtained by the definition of smooth Rényi zero-entropy (\ref{smoothRenyizero}):
		\begin{eqnarray}
		&\fl \forall i\in\{1,\dots,N-1\} \quad \exists \bar{P}_{XK_i} \quad\mathrm{s.t.}  \quad \nonumber \\
		&\fl\norm{\bar{P}_{XK_i}-P_{XK_i}} \leq 2\varepsilon' \quad \wedge \quad H_{0}(\bar{P}_{XK_i}|K_i) = H_{0}^{\varepsilon'}(P_{XK_i}|K_i) \,\,, \label{Pbar}
		\end{eqnarray}
		where the distance between two probability distributions is defined as: $$\norm{P-Q} = \sum_x |P(x) - Q(x)|\,\,.$$
		We define $\bar{i} \equiv \arg\max_i H_{0}(\bar{P}_{XK_i}|K_i)$, then (\ref{Pbar}) implies:
		\begin{eqnarray}
		\max_i H_{0}(\bar{P}_{XK_i}|K_i) = H_{0}(\bar{P}_{XK_{\bar{i}}}|K_{\bar{i}}) 
		&= H_{0}^{\varepsilon'}(P_{XK_{\bar{i}}}|K_{\bar{i}})  \nonumber \\
		&\leq \max_i H_{0}^{\varepsilon'}(P_{XK_i}|K_i) \label{Eqn1} \,\,.
		\end{eqnarray}
		Let us now consider the protocol $\mathrm{EC}_{\hat{\mathcal{X}},\mathcal{F}}$ where $\hat{\mathcal{X}}$ and $\mathcal{F}$ are fixed by the above-defined
		set of distributions $\set{\bar{P}_{XK_i}}_{i=1}^{N-1}$. Then, by (\ref{intermediate-result}) we know that such an EC protocol is 
		$\varepsilon_{\mathrm{EC}}$-fully secure and has leakage:
		\begin{eqnarray}
		\mathrm{leak}_{\mathrm{EC}_{\hat{\mathcal{X}},\mathcal{F}}}^{\mathrm{NQKD}}
		&\leq \max_i H_{0}(\bar{P}_{XK_i}|K_i) + \log_2 \left(2/\varepsilon_{\mathrm{EC}}\right) + \log_2 (N-1) \nonumber \\
		&\leq \max_i H_{0}^{\varepsilon'}(P_{XK_i}|K_i) + \log_2 \left(2/\varepsilon_{\mathrm{EC}}\right) + \log_2 (N-1) \nonumber 
		\end{eqnarray}
		where we used (\ref{Eqn1}) in the second inequality.\\
		The last thing to be shown is that such an EC protocol is also $2(N-1)\varepsilon'$-robust on the distribution $P_{X\mathbf{K}}$:
		\begin{equation}
		\mathrm{Pr}_{(x,\mathbf{k})} [\mathrm{abort}]_{P}\leq 2(N-1)\varepsilon' \,\,, \nonumber 
		\end{equation}
		i.e. the probability that the protocol aborts when initiated with a set of keys $(x,\mathbf{k})$ generated by the distribution $P_{X\mathbf{K}}$ is lower or equal
		than $2(N-1)\varepsilon'$\footnote{Note that this EC protocol is defined by the distributions $\bar{P}_{XK_i}$ which are one by one 
		$2\varepsilon'$-close to the marginals of the distribution $P_{X\mathbf{K}}$ defining the EC protocol of part 1, which was shown to be $0$-robust on 
		$P_{X\mathbf{K}}$. It is not straightforward to infer -unlike the bipartite case- that the new EC protocol is then $(N-1)\cdot 2\varepsilon'$-robust on
		$P_{X\mathbf{K}}$.}. Let us compute the probability of $\mathrm{EC}_{\hat{\mathcal{X}},\mathcal{F}}$ to abort:
		\begin{eqnarray}
		\mathrm{Pr}_{(x,\mathbf{k})} [\mathrm{abort}]_{P} &= \mathrm{Pr}_{(x,\mathbf{k})}\left[\exists\,i : \hat{\mathcal{D}}_i =\emptyset \right]_{P}  \nonumber \\
		&= 1-  \mathrm{Pr}_{(x,\mathbf{k})}\left[\hat{\mathcal{D}}_i \neq\emptyset \,\forall i \right]_{P} \,\,. \nonumber
		\end{eqnarray}
		One of the possibilities for $\hat{\mathcal{D}}_i$ not to be empty is $x\in\hat{\mathcal{D}}_i \,\Leftrightarrow\,x\in \hat{\mathcal{X}}_{k_i}^i \,\Leftrightarrow
		\bar{P}_{XK_i}(x,k_i)\neq 0$, which is not obvious since $x$ was generated through the distribution $P_{X\mathbf{K}}$. Therefore:
		\begin{equation}
		\mathrm{Pr}_{(x,\mathbf{k})}\left[\hat{\mathcal{D}}_i \neq\emptyset \,\forall i \right]_{P} \geq 
		\mathrm{Pr}_{(x,\mathbf{k})}\left[\bar{P}_{XK_i}(x,k_i)\neq 0 \,\forall i \right]_{P}  \,\,. \label{Eqn2}
		\end{equation}
		By employing the following inequality from probability theory (straightforward proof based on union bound and de-Morgan's law):
		\begin{equation}
		\mathrm{Pr}\left(\bigcap_{i=1}^n A_i\right) \geq \sum_{i=1}^n \mathrm{Pr}(A_i) -(n-1) \label{probability-law}
		\end{equation}
		where $\mathrm{Pr}(A_i)$ is the probability of event $A_i$,	we are able to recast the r.h.s. of (\ref{Eqn2}) as:
		\begin{eqnarray}\fl
		\mathrm{Pr}_{(x,\mathbf{k})}\left[\hat{\mathcal{D}}_i \neq\emptyset \,\forall i \right]_{P} &\geq 
		\mathrm{Pr}_{(x,\mathbf{k})}\left[\bar{P}_{XK_i}(x,k_i)\neq 0 \,\forall i \right]_{P}  \nonumber \\
		&\geq \sum_{i=1}^{N-1} \mathrm{Pr}_{(x,k_i)}\left[\bar{P}_{XK_i}(x,k_i)\neq 0\right]_{P} - [(N-1)-1] \,\,.  \label{Eqn3}
		\end{eqnarray}
		We now concentrate on computing $\mathrm{Pr}_{(x,k_i)}\left[\bar{P}_{XK_i}(x,k_i)\neq 0\right]_{P}$, which is the probability that, having generated the couple 
		$(x,k_i)$ from distribution $P_{XK_i}$, it holds that $\bar{P}_{XK_i}(x,k_i)\neq 0$. We employ the fact that by assumption (\ref{Pbar}) 
		the distance between the two involved distributions is bounded by $2\varepsilon'$, which implies that, for instance:
		\begin{equation}
		\left\lvert P_{XK_i}(x,k_i) - \bar{P}_{XK_i}(x,k_i) \right\rvert \leq 2\varepsilon' \quad \forall \,(x,k_i)  \,\,.\label{Eqn4}
		\end{equation}
		Let us focus on the probability of the complementary event: $\mathrm{Pr}_{(x,k_i)}\left[\bar{P}_{XK_i}(x,k_i)= 0\right]_{P}$. Since this event is a sufficient 
		condition for having $P_{XK_i}(x,k_i) \leq 2\varepsilon'$ (because of (\ref{Eqn4})), this means that:
		\begin{equation}
		\mathrm{Pr}_{(x,k_i)} \left[P_{XK_i}(x,k_i) \leq 2\varepsilon'\right]_P \geq \mathrm{Pr}_{(x,k_i)} \left[\bar{P}_{XK_i}(x,k_i)= 0\right]_P \,\,,\label{Eqn5}
		\end{equation}
		but the l.h.s of (\ref{Eqn5}) can be bounded by:
		\begin{equation}
		\mathrm{Pr}_{(x,k_i)} \left[P_{XK_i}(x,k_i) \leq 2\varepsilon'\right]_P \leq 2\varepsilon' \,\,, \nonumber
		\end{equation}
		therefore we have:
		\begin{equation}
		\mathrm{Pr}_{(x,k_i)}\left[\bar{P}_{XK_i}(x,k_i)\neq 0\right]_{P} \geq 1- 2\varepsilon' \,\,. \label{Eqn6}
		\end{equation}
		Substituting in (\ref{Eqn3}) yields:
		\begin{equation}\fl
		\mathrm{Pr}_{(x,\mathbf{k})}\left[\hat{\mathcal{D}}_i \neq\emptyset \,\forall i \right]_{P} \geq 
		(N-1)(1-2\varepsilon') +1 -(N-1) = 1- (N-1)2\varepsilon' \,\,.\label{Eqn7}  
		\end{equation}
		With this result we can conclude that:
		\begin{eqnarray}
		\mathrm{Pr}_{(x,\mathbf{k})} [\mathrm{abort}]_{P} &= 1-  \mathrm{Pr}_{(x,\mathbf{k})}\left[\hat{\mathcal{D}}_i \neq\emptyset \,\forall i \right]_{P} \nonumber \\
		&\leq 2(N-1)\varepsilon' \nonumber
		\end{eqnarray}
		which concludes the proof.\hfill\opensquare
	
	\section{Quantifying the channel's noise} \label{PE}
	As anticipated in \autoref{NQKD-protocols}, one can bound $E$'s knowledge about the secret key by quantifying the noise she introduced in the quantum channel.\\
	In this Section we show how the relevant noise parameters of both protocols can be estimated from the finite statistics collected in PE.
	\subsection{N-BB84 protocol}
	In the $N$-BB84 protocol, the important noise parameters that are subsequently used to characterize $E$'s knowledge are $Q_{A B_i}^n$ and $Q_X^n$, i.e. the
	frequency of discordant $Z$-outcomes between $A$ and $B_i$ and the frequency of the outcome $X^{\otimes N}=-1$, respectively. Both frequencies refer to hypothetical
	measurements performed on the remaining $n$ signals following PE. The goal is to characterize the noise parameters based on what is observed in PE 
	($Q_{A B_i}^m$ and $Q_X^m$). This is easily achieved by means of the following Lemma (generalization of a result presented in \cite[Suppl. Note 2]{TomamichelNature}):
	\begin{lemma} \label{thm-PE}
		Let $\varepsilon>0$. Let $\mathbf{R}$ be a random binary string of $M=n+m$ bits with relative Hamming weight $\Lambda_{M}=\frac{1}{M}|\mathbf{R}|$. 
		Let $R_1,\dots,R_m$ be random variables obtained by sampling $m$ random entries of $\mathbf{R}$ without replacement. Then, upon defining:
		\begin{eqnarray}
		\Lambda_m &= \frac{\sum_{i=1}^m R_i}{m}= \frac{|(\mathbf{R})_m|}{m}  \label{Lambda-m} \\
		\Lambda_n &= \frac{|(\mathbf{R})_n|}{n}  \label{Lambda-n}
		\end{eqnarray}
		as the relative Hamming weights\footnote{We denote by $(\mathbf{R})_m$ the $m$-bit string composed by the random variables $R_1,\dots,R_m$, while $(\mathbf{R})_n$
		is the $n$-bit string composed by the remaining entries of $\mathbf{R}$.} of the two randomly chosen partitions of $\mathbf{R}$, it holds:
		\begin{eqnarray}
		\Pr\left[\frac{1}{2} |\Lambda_n - \Lambda_m| > \xi(\varepsilon,n,m)\right] &\leq 2\varepsilon  \nonumber \\
		\Pr\left[\Lambda_n > \Lambda_m + 2\xi(\varepsilon,n,m)\right] &\leq \varepsilon  \nonumber \\
		\Pr\left[\Lambda_m > \Lambda_n + 2\xi(\varepsilon,m,n)\right] &\leq \varepsilon   \label{Th-prob}  
		\end{eqnarray}
		where:
		\begin{equation}
		\xi(\varepsilon,n,m) \equiv \sqrt{\frac{(n+m)(m+1)}{8nm^2} \ln \left(\frac{1}{\varepsilon}\right)}  \,\,.  \label{Th-xi}
		\end{equation}
	\end{lemma}
	\textit{Proof}.	
	Let's first fix the random bit string $\mathbf{R}$ to a given and known string: $\mathbf{R}\equiv\mathbf{r}$; thus also its relative Hamming weight is fixed to
	some real value: $\Lambda_M \equiv \lambda_M$. Then it holds \cite[Theorem 1]{BoumanFehr}:
	\begin{eqnarray}
	\Pr \left[|\Lambda_n - \lambda_M| > \delta\,\, \bigg\rvert\,\, \mathbf{R}=\mathbf{r}, \Lambda_M = \lambda_M\right] &\leq 2\, e^{-2\frac{nM}{m+1}\delta^2}
	\label{thm1.1} \\
	\Pr \left[\Lambda_n > \lambda_M + \delta\,\, \bigg\rvert\,\, \mathbf{R}=\mathbf{r}, \Lambda_M = \lambda_M\right] &\leq e^{-2\frac{nM}{m+1}\delta^2}
	\,\,. \label{thm1.2}
	\end{eqnarray}
	By defining $\nu=\frac{m}{M}$, it is immediate to show the following facts for every $\mu\in\mathbb{R}$:
	\begin{eqnarray}
	&\Lambda_M = \nu \Lambda_m + (1-\nu) \Lambda_n   \nonumber  \\
	&|\Lambda_n - \Lambda_M| > \nu\mu \quad\iff\quad |\Lambda_n - \Lambda_m| > \mu  \label{thm2} \\
	&\Lambda_n > \Lambda_M + \nu\mu \quad\iff\quad \Lambda_n > \Lambda_m + \mu  \,\,. \label{thm3} 
	\end{eqnarray}
	Now one can make use of (\ref{thm1.1}) and (\ref{thm2}) in the following calculation:
	\begin{eqnarray}
	\fl\Pr\left[|\Lambda_n - \Lambda_m| > \mu \right] &=\Pr\left[|\Lambda_n - \Lambda_M| > \nu\mu\right]  \nonumber \\
	& =\sum_{\mathbf{r}}\Pr\left[\mathbf{R}=\mathbf{r}\right]\Pr \left[|\Lambda_n - \lambda_M| > \nu\mu\,\, \bigg\rvert\,\,
	\mathbf{R}=\mathbf{r}, \Lambda_N = \lambda_M\right]  \nonumber \\
	&\leq \sum_{\mathbf{r}}\Pr\left[\mathbf{R}=\mathbf{r}\right] 2 \,e^{-2\frac{nM}{m+1}\frac{m^2}{M^2}\mu^2} \nonumber \\
	&= 2\,e^{-2\frac{nm^2}{(m+1)M}\mu^2}  \,\,.  \label{thm4}
	\end{eqnarray}
	Analogously, by using (\ref{thm1.2}) and (\ref{thm3}) one obtains:
	\begin{equation}
	\Pr\left[\Lambda_n > \Lambda_m + \mu \right] \leq e^{-2\frac{nm^2}{(m+1)M}\mu^2}  \,\,.  \label{thm5}
	\end{equation}
	Finally, by choosing $\mu$ such that it holds: $e^{-2\frac{nm^2}{(m+1)M}\mu^2} = \varepsilon$, i.e. $\mu= 2\xi(\varepsilon,n,m)$ with $\xi(\varepsilon,n,m)$ 
	defined as in (\ref{Th-xi}), one obtains from (\ref{thm4}) and (\ref{thm5}):
	\begin{eqnarray}
	\Pr\left[\frac{1}{2}|\Lambda_n - \Lambda_m| > \xi(\varepsilon,n,m) \right] &\leq 2\varepsilon  \nonumber \\
	\Pr\left[\Lambda_n > \Lambda_m + 2\xi(\varepsilon,n,m)\right] &\leq \varepsilon  \nonumber
	\end{eqnarray}
	which is exactly the claimed result in (\ref{Th-prob}). The last expression in (\ref{Th-prob}) is simply obtained by exchanging the roles of $n$ and
	$m$.\hfill\opensquare\bigskip\\
	In order to make use of \autoref{thm-PE}, we define the following random vectors containing the outcomes of $A$ and $B_i$'s $Z$-measurement rounds devoted to PE:
	\begin{eqnarray}
	(\mathbf{Z_a})_j \equiv \left\{  \begin{array}{r@{\quad}cr} 
	1 & z_{a,j}=-1 \\
	0 & z_{a,j}=1   
	\end{array}\right.
	\quad 
	(\mathbf{Z_i})_j \equiv \left\{  \begin{array}{r@{\quad}cr}
	1 & z_{i,j}=-1 \\
	0 & z_{i,j}=1 
	\end{array}\right.    \,\,.
	\end{eqnarray}
	Analogously, we define the random vectors containing the outcomes of $A$ and \textbf{B}'s $X$-measurement rounds:
	\begin{eqnarray}
	(\mathbf{X_a})_j \equiv  \left\{  \begin{array}{r@{\quad}cr}
	1 & x_{a,j}=-1 \\
	0 & x_{a,j}=1 
	\end{array}\right.
	\quad 
	(\mathbf{X_i})_j \equiv \left\{  \begin{array}{r@{\quad}cr}
	1 & x_{i,j}=-1 \\
	0 & x_{i,j}=1 
	\end{array}\right.   \,\,. \label{X}
	\end{eqnarray}
	With these definitions, it holds:
	\begin{equation}
	(\mathbf{X_a}\oplus\mathbf{X_1}\oplus\dots\oplus\mathbf{X_{N-1}})_j =  \left\{  \begin{array}{r@{\quad}cr}
	1 & \left(x_a \prod_{i=1}^{N-1} x_i\right)_j= -1 \\
	0 & \left(x_a \prod_{i=1}^{N-1} x_i\right)_j=1 
	\end{array}\right.  \,\,  
	\end{equation}
	therefore it is immediate to verify that:
	\begin{eqnarray}
	Q^m_{A B_i} &= \frac{|\mathbf{Z_a}\oplus \mathbf{Z_i}|}{m}  \nonumber \\
	Q^m_X &= \frac{|\mathbf{X_a}\oplus\mathbf{X_1}\oplus\dots\oplus\mathbf{X_{N-1}}|}{m} \label{Q_X} \,\,.
	\end{eqnarray}
	Since we were able to write the frequencies $Q_{A B_i}^m$ and $Q_X^m$ as relative Hamming weights of random vectors, we can apply \autoref{thm-PE} and state that:
	\begin{eqnarray}\fl
	\Pr\left[Q^n_{X} \leq Q^m_{X} + 2\xi(\varepsilon_x,n,m) \,\,\wedge \,\, Q^n_{A B_i} \leq Q^m_{A B_i} + 2\xi(\varepsilon_z,n,m) \,\,\forall i \right] 
	\geq 1- \varepsilon^2_{\mathrm{PE}}  \label{probALL2} 
	\end{eqnarray}
	where we used (\ref{probability-law}) and defined:
	\begin{equation}
	\varepsilon_{\mathrm{PE}} \equiv \sqrt{(N-1)\varepsilon_z + \varepsilon_x}  \label{epsilon-PE-2} \,\,.
	\end{equation}
		
	\subsection{N-six-state protocol}
	In this case $E$ is supposed to gain information about the key only via collective attacks, i.e. she attacks each of the shared signals independently and
	identically\footnote[1]{Then the result is extended to coherent attacks via the PS technique, see \ref{details-computablerates} for the details.}. Thus, 
	the needed noise parameters are the probabilities $P_X$,$P_{A B_i}$ and $P_Z$ computed on a single
    $N$-qubit signal state, which in turn has a very simple expression \cite[Eq. 11]{Epping} thanks to the extended depolarization procedure.\\
    The PE frequencies $Q^{m'}_X$, $Q^m_{A B_i}$ and $Q_Z^m$ are thus observed on multiple copies of the same $N$-qubit signal state. Therefore they constitute an 
    estimation of the corresponding probabilities by the Law of Large Numbers \cite{Mertz2011}:
    \begin{eqnarray} \fl
    \Pr\left[\frac{1}{2} |Q_{AB_i}^m - P_{AB_i}| \leq \eta(\varepsilon_z,2,m)\,\,\forall\, i \,\,\wedge\,\, \frac{1}{2} |Q^{m'}_X - P_X| \leq\eta(\varepsilon_x,2,m')
    \right. \nonumber \\ \left.\wedge\,\,\frac{1}{2}|Q_{Z}^m - P_Z| \leq \eta(\varepsilon_z',2,m) \right]
    \geq 1-\varepsilon_{\mathrm{PE}}  \label{probALL-Protocol1}
    \end{eqnarray}
    where we used (\ref{probability-law}) and defined:
    \begin{eqnarray}
    \varepsilon_{\mathrm{PE}} \equiv \varepsilon_z' + (N-1)\varepsilon_z +\varepsilon_x   \label{epsilon-PE-1}  \\\vspace{1pt}
    \eta(\varepsilon,d,m) \equiv \sqrt{\frac{\ln (1/\varepsilon)+d\ln (m+1)}{8 m}} \quad. \label{eta}
    \end{eqnarray}
	
	\section{Derivation of the computable key lengths}  \label{details-computablerates}
    In order to obtain a computable key length for the $N$-BB84 (\autoref{computable-length-NBB84}) and the $N$-six-state (\autoref{computable-length-Nsixstate}) protocol
    starting from the general result (\autoref{achievable1and2way}), one needs to lower bound the min-entropy (which quantifies
    $E$'s uncertainty about the key) and to upper bound the leakage term with quantities depending on the channel's noise.\\
    In this Section we show how to achieve this task for both protocols and how to further characterize the noise via the PE finite statistics, by using the results of
    \ref{PE}. \\
    Concerning the notation, for the remainder of the Section we indicate with an apex the number of signals described by the quantum state, and we also indicate as
    $Z$ the classical system containing $A$'s raw key bits (since in both protocols the raw keys are generated by $Z$-basis measurements). Thus the quantum state 
    describing the parties' raw keys and $E$'s degree of freedom is indicated as: $\rho^n_{Z\mathbf{K}E}$. 
    
    \subsection{N-BB84 protocol}
    \textbf{Leakage.} 
    The leakage of an optimal 1-way EC protocol (\ref{leak-upperbound}) is bounded by the smooth Rényi zero-entropy of the probability distribution of $A$ and $B_i$'s 
    raw keys (\ref{smoothRenyizero}). Note that, thanks to (\ref{Bpurified-in-B}), we can bound such an entropy by:
    \begin{equation}
    H_{0}^{\varepsilon_{\mathrm{PE}}}(P_{ZK_i}^n|K_i) \leq H_{0}^{\varepsilon_{\mathrm{PE}},\,\mathrm{P}}(P_{ZK_i}^n|K_i) 
    \label{step1-Protocol2}   \,\,.
    \end{equation}
    In this way, one can follow the proof of \cite[Lemma 3]{TomamichelNature} and show that there exists a probability
    distribution $R_{ZK_i}^n\in \mathcal{B}^{\varepsilon_{\mathrm{PE}},\,\mathrm{P}} (P_{ZK_i}^n)$ such that the frequency of discordant bits ($Q^n_{A B_i}$)
    is less or equal than $Q^m_{A B_i} + 2\xi(\varepsilon_z,n,m)$, with certainty. Note that this is not true for the distribution $P_{ZK_i}^n$, since
    it holds condition (\ref{probALL2}).\\
    This upper limit on the number of discordant bits between $A$ and $B_i$, when the keys are generated by $R_{ZK_i}^n$, allows one to bound the 
    Rényi zero-entropy of such a distribution by $n h\left(Q_{AB_i}^m +2\xi(\varepsilon_z,n,m)\right)$.\\
    Finally, since the smooth Rényi entropy of order zero is defined with a \emph{minimization} over its $\varepsilon$-environment
    (\ref{smoothRenyizero}), one obtains: 
    \begin{equation}
    H_{0}^{\varepsilon_{\mathrm{PE}},\,\mathrm{P}}(P_{ZK_i}^n|K_i) \leq n h\left(Q_{AB_i}^m +2\xi(\varepsilon_z,n,m)\right)
    \label{step2-Protocol2}   \,\,.
    \end{equation}
    Combining (\ref{step1-Protocol2}) and (\ref{step2-Protocol2}) with \autoref{leak_upperbound} leads to the desired result. The leakage occurring in the $N$-BB84 
    protocol, implemented with the optimal 1-way, $\varepsilon_{\mathrm{EC}}$-fully secure and $2(N-1)\varepsilon_{\mathrm{PE}}\,$-robust EC protocol, is:
    \begin{equation}
    \mathrm{leak}_{\mathrm{EC}}^{\mathrm{NQKD}} \leq n \max_i h\left(Q_{AB_i}^m +2\xi(\varepsilon_z,n,m)\right) + 
    \log_2 \frac{2(N-1)}{\varepsilon_{\mathrm{EC}}} \,\,.  \label{leak-NBB84}
    \end{equation}
    
    \textbf{Min-entropy.}
    Let $\rho^{n+2m}_{A\mathbf{B}E}$ be the pure state describing the whole set of quantum signals and $E$'s quantum system. The state $\rho^n_{ZE}$ is then obtained by
    performing independent $Z$-measurements on $A$'s subsystems and taking the partial trace over \textbf{B}'s ones, after the PE procedure took place on $2m$ signals.
    If we now define $\rho^n_{X\mathbf{B}}$ as the state obtained by performing independent $X$-measurements on $A$'s subsystems and then taking the partial trace over $E$,
    we can employ the uncertainty relation \cite{uncertainty-rel}:
    \begin{equation}
    H_{\mathrm{min}}^{\bar{\varepsilon},\, \mathrm{P}}\left(\rho_{ZE}^n|E\right) \geq q - H_{\mathrm{max}}^{\bar{\varepsilon},\, \mathrm{P}}
    \left(\rho_{X\mathbf{B}}^n|\mathbf{B}\right)  \label{uncert-rel}
    \end{equation}
    where $q=-\log_2 c$, with:
    \begin{equation}
    c = \max_{\mathbf{z},\mathbf{x}} \norm{(P_{z_1}\otimes\dots\otimes P_{z_n})(P_{x_1}\otimes\dots\otimes P_{x_n})}^2_{\infty}  \nonumber
    \end{equation}
    and $P_{z_1}\otimes\dots\otimes P_{z_n}$, $P_{x_1}\otimes\dots\otimes P_{x_n}$ are the projectors implementing the $Z$- and $X$-measurements on $A$'s 
    subsystems, respectively.
    In particular, $P_{z_i}\in\{P_{\ket{0}},P_{\ket{1}}\}$ and $P_{x_i}\in\{P_{\ket{+}},P_{\ket{-}}\}$. Therefore one can easily compute the quality factor $q$ in this
    specific case: $q=n$\footnote{The norm $\norm{\cdot}_{\infty}$ evaluates the largest singular value.}.\\
    We can now bound the max-entropy (\ref{smooth-max-entropy}) of the classical-quantum states $\rho_{X\mathbf{B}}^n$ by performing the same projective measurement on
    all \textbf{B}'s subsystems and by employing the data processing inequality \cite[Theorem 6.2]{TomamichelBook}:
    \begin{equation}
    H_{\mathrm{max}}^{\bar{\varepsilon},\, \mathrm{P}} \left(\rho_{X\mathbf{B}}^n|\mathbf{B}\right) \leq \
    H_{\mathrm{max}}^{\bar{\varepsilon},\, \mathrm{P}} \left(\rho_{X\mathbf{X}}^n|\mathbf{X}\right)   \nonumber
    \end{equation}
    which inserted in (\ref{uncert-rel}) yields:
    \begin{equation}
    H_{\mathrm{min}}^{\bar{\varepsilon},\, \mathrm{P}}\left(\rho_{ZE}^n|E\right) \geq n - H_{\mathrm{max}}^{\bar{\varepsilon},\, \mathrm{P}}
    \left(\rho_{X\mathbf{X}}^n|\mathbf{X}\right)  \label{uncert-rel-new}  \,\,.
    \end{equation}
    Finally one can bound the max-entropy of the classical state $\rho_{X\mathbf{X}}^n$, - i.e. of the probability distribution $P^n_{X\mathbf{X}}$ -
    by means of \cite[Lemma 3]{TomamichelNature}. As a matter of fact, one can consider the whole set of \textbf{B} as one single Bob with the $X$-outcomes vector
    defined as:
    \begin{equation}
    \mathbf{X}'= \mathbf{X_1}\oplus\dots\oplus\mathbf{X_{N-1}}   \label{X'} \,\,,
    \end{equation}
    where the random vectors are defined in (\ref{X}). Under this classical operation the data processing inequality holds:
    \begin{equation}
    H_{\mathrm{max}}^{\bar{\varepsilon},\, \mathrm{P}} \left(\rho_{X\mathbf{X}}^n|\mathbf{X}\right) = 
    H_{\mathrm{max}}^{\bar{\varepsilon},\, \mathrm{P}} \left(P_{X\mathbf{X}}^n|\mathbf{X}\right) \leq 
    H_{\mathrm{max}}^{\bar{\varepsilon},\, \mathrm{P}} \left(P_{XX'}^n|X'\right)   \,\,. \label{Bobs-as-one}
    \end{equation}
    In this fashion, the PE parameter $Q_X^m$ is exactly the frequency of discordant bits between $\mathbf{X_a}$ and $\mathbf{X}'$ (see its definition in (\ref{Q_X})). 
    Therefore one can apply \cite[Lemma 3]{TomamichelNature}:
    \begin{equation}
    H_{\mathrm{max}}^{\varepsilon_{\mathrm{PE}},\, \mathrm{P}} \left(P_{XX'}^n|X'\right) \leq n h\left(Q^m_X + 2\xi(\varepsilon_x,n,m)\right)  \nonumber
    \end{equation}
    which combined with (\ref{Bobs-as-one}) yields:
    \begin{equation}
    H_{\mathrm{max}}^{\varepsilon_{\mathrm{PE}},\, \mathrm{P}} \left(\rho_{X\mathbf{X}}^n|\mathbf{X}\right) \leq n h\left(Q^m_X + 2\xi(\varepsilon_x,n,m)\right)
    \label{Lemma3-for-min-entropy}  \,\,.
    \end{equation}
    Finally inserting (\ref{Lemma3-for-min-entropy}) in (\ref{uncert-rel-new}) after having fixed: $\bar{\varepsilon}=\varepsilon_{\mathrm{PE}}$, yields the desired 
    result:
    \begin{equation}
    H_{\mathrm{min}}^{\varepsilon_{\mathrm{PE}},\, \mathrm{P}}\left(\rho_{ZE}^n|E\right) \geq n(1 - h\left(Q^m_X + 2\xi(\varepsilon_x,n,m)\right))
    \label{min-entropy-NBB84}  \,\,.
    \end{equation}
    \textbf{Computable key length.} By employing the bounds on the leakage (\ref{leak-NBB84}) and on the min-entropy (\ref{min-entropy-NBB84})
    in \autoref{achievable1and2way}, one obtains the computable key length presented in \autoref{computable-length-NBB84}, which only depends on the PE statistics and 
    on the security parameters.
       
    \subsection{N-six-state protocol}
    As anticipated in \autoref{computablerates}, the strategy adopted to achieve a computable expression of the $N$-six-state key length relies on the PS technique
    \cite{postsel}.
    Such a technique allows to prove a given property of a quantum channel, acting on a general multipartite state, by just proving it on inputs consisting of
    identical and independent copies of a state on a single subsystem. 
    Therefore one can infer the security of a QKD protocol -viewed as a quantum channel- under coherent attacks (arbitrary input)
    from the security of the same protocol under collective attacks (product state input) \cite{Sheridan}.
    For this reason in the following we restrict $E$'s action to collective attacks, meaning that the quantum state describing the parties' raw keys and $E$'s quantum 
    system is a product state: $\rho^{\otimes n}_{Z\mathbf{K}E}$, and the raw keys' probability distribution is a product distribution: $(P_{Z\mathbf{K}})^n$.\\
    \textbf{Leakage.}
    We start from the general upper bound stated in (\ref{leak-upperbound}) and employ the finite version of the AEP for probability distributions 
    \cite[Theorem 1]{non-asymptotic-AEP} to further bound the smooth Rényi zero-entropy (\ref{smoothRenyizero}):
    \begin{eqnarray}
    H^{\varepsilon_{\mathrm{PE}}}_0((P_{ZK_i})^n|K_i) \leq n \left[H(Z|K_i) + \log_2 (5) \, \sqrt{\frac{2\log_2 (1/(2\varepsilon_{\mathrm{PE}}))}{n}} \right]
    \label{leak1}
    \end{eqnarray}
    where we fixed $\varepsilon'=\varepsilon_{\mathrm{PE}}$ as defined in (\ref{epsilon-PE-1}) and where $H(Z|K_i)$ is the conditional Shannon entropy of $P_{Z K_i}$. 
    Thanks to the symmetries introduced by the extended depolarization procedure \cite{Epping} each raw key bit is uniform: $H(Z)=H(K_i)=1$. These constraints on the 
    probability distribution $P_{Z K_i}$ imply that its conditional entropy $H(Z|K_i)$ can be expressed as a function of the only parameter
    $P_{A B_i}$ as follows: $H(Z|K_i)=h(P_{A B_i})$.\\
    Finally, we characterize the probability $P_{A B_i}$ through the observed frequency $Q^m_{AB_i}$ in PE (\ref{probALL-Protocol1}). In particular, we exploit the 
    composable-security property by adding $\varepsilon_{\mathrm{PE}}$ to the total security parameter and by maximizing (\ref{leak1}) over the allowed probabilities.
    Combining this with \autoref{leak_upperbound} leads to the desired result. The leakage occurring in the $N$-six-state 
    protocol, implemented with the optimal 1-way, $\varepsilon_{\mathrm{EC}}$-fully secure and $2(N-1)\varepsilon_{\mathrm{PE}}\,$-robust EC protocol, is:
   	\begin{eqnarray}
    	\fl\mathrm{leak}_{\mathrm{EC}}^{\mathrm{NQKD}} \leq n \left[\max_i h\left(Q_{AB_i}^m +2\eta(\varepsilon_z,2,m)\right) 
    	+\log_2 (5) \, \sqrt{\frac{2\log_2 (1/(2\varepsilon_{\mathrm{PE}}))}{n}}\right] \nonumber \\
    	+ \log_2 \frac{2(N-1)}{\varepsilon_{\mathrm{EC}}} \,\,.  \label{leak-Nsixstate}
   	\end{eqnarray}
    
    \textbf{Min-entropy.}
    We can bound the min-entropy of a product state via the finite version of the AEP for quantum states, reported in \cite[Equation B7]{Mertz2013}:
    \begin{eqnarray}
    H_{\mathrm{min}}^{\bar{\varepsilon}} (\rho^{\otimes n}_{ZE} |E)  \geq n \left(S(\rho_{ZE}) - S(\rho_E) - 
    5\sqrt{\frac{\log_2 (1/\bar{\varepsilon})}{n}}\right)  \label{AEP}  \,\,,
    \end{eqnarray}
    where $S(\rho)$ is the Von Neumann entropy. The r.h.s. of (\ref{AEP}) can be recast in terms of the probabilities $P_X$ and $P_Z$, by following analogous steps
    in \cite{Epping} and by exploiting the symmetries of the single-signal state due to the extended depolarization procedure. \\
    Finally, the probabilities $P_X$ and $P_Z$ are characterized by the PE measurements through (\ref{probALL-Protocol1}). Thus we can minimize the min-entropy bound 
    over the allowed probabilities while adding the PE failure probability $\varepsilon_{\mathrm{PE}}$ to the total security parameter. These operations yield:
    \begin{eqnarray}
    \fl H_{\mathrm{min}}^{\bar{\varepsilon}} (\rho^{\otimes n}_{ZE}|E)  \geq n  \inf_{\Gamma_{\mathrm{PE}}} 
    \left[\left(1-\frac{P_Z}{2} -P_X\right)\log_2 \left(1-\frac{P_Z}{2}
    -P_X\right) \right. \nonumber \\
    \fl \left. + \left(P_X -\frac{P_Z}{2}\right) \log_2 \left(P_X -\frac{P_Z}{2}\right) + (1-P_Z) \left(1-\log_2 (1-P_Z)\right) - 
    5\sqrt{\frac{\log_2 (1/\bar{\varepsilon})}{n}}\right] \label{min-entropy-Nsixstate}
    \end{eqnarray}
    where the set $\Gamma_{\mathrm{PE}}$ is defined in (\ref{GammaPE}).\\
    \textbf{Computable key length.} By substituting the bounds (\ref{leak-Nsixstate}) and (\ref{min-entropy-Nsixstate}) into \autoref{achievable1and2way}, one obtains
    the computable key length of the $N$-six-state protocol when performed under collective attacks. \\
    The PS technique \cite{postsel} allows to extend the security of a protocol against collective attacks, to any kind of attack, by just shortening the key length and 
    introducing a corrective factor on the total security parameter.
    Consider an NQKD protocol $\mathcal{E}$ acting on $L$-partite systems (the $L$ shared signals), 
    where each of the $L$ constituents has dimension $d$ (in our case each signal describes the state of $N$ qubits, thus $d=2^N$). 
    If $\mathcal{E}$ is $\varepsilon_{\mathrm{tot}}$-secure against collective attacks, then the protocol $\mathcal{E}'$ obtained from $\mathcal{E}$ by shortening 
    the output of the hashing by ``$2(d^2-1)\log_2 (L+1)$'' bits is $(L+1)^{(d^2-1)}\varepsilon_{\mathrm{tot}}$-secure against coherent attacks.
    By applying the PS corrections to the $N$-six-state key valid for collective attacks, we extend its validity to coherent attacks, yielding the final result: 
    \autoref{computable-length-Nsixstate}.
    
    \section{Information leaked from the classical channel}  \label{RennerThm}
    The following Lemma is the result of a private communication \cite{privateComm} with Renato Renner. 
    It shows that the additional information that $E$ has about $A$'s raw
    key $X$ due to EC's classical communication can be quantified by the leakage (as defined in \autoref{def-leak}), even for a general two-way EC protocol. The proof
    relies on the fact that the \mbox{$\varepsilon$-environment} of the entropies is defined via the purified distance. 
    The crucial advantage of this definition of distance is
    that one can always find extensions and purifications of quantum states without increasing their distance \cite{purifieddistanceAdvantage}.
    \begin{lemma}
    	Let $\rho_{X\mathbf{KC}E}$ be a density operator with $X,\mathbf{K},\mathbf{C}$ classical, such that the Markov chain condition $\mathbf{C}\leftrightarrow
    	(X,\mathbf{K})\leftrightarrow E$ holds. Then, for any $\varepsilon \geq 0$,
    	\begin{equation}
    	 H_{\mathrm{min}}^{\varepsilon,\, \mathrm{P}}(\rho_{X \mathbf{C}E}|\mathbf{C}E) \geq H_{\mathrm{min}}^{\varepsilon,\,\mathrm{P}}(\rho_{X E}|E)  
    	 - H_{0}(\rho_{\mathbf{C}}) + H_{\mathrm{min}}(\rho_{X\mathbf{K}\mathbf{C}}|\rho_{X\mathbf{K}})  \,\,. \label{RennerLemma}	
    	\end{equation}
    \end{lemma}
    \textit{Proof}. We first prove the statement in the special case where $\varepsilon=0$. This is achieved by the following chain of inequalities:
    \begin{eqnarray}
    	H_{\mathrm{min}}(\rho_{X \mathbf{C}E}|\mathbf{C}E) &\stackrel{(1)}{\geq} H_{\mathrm{min}}(\rho_{X \mathbf{C}E}|E) - H_{0}(\rho_{\mathbf{C}}) \nonumber\\
    	&\stackrel{(2)}{\geq} H_{\mathrm{min}}(\rho_{X \mathbf{C}E}|\rho_{XE}) + H_{\mathrm{min}}(\rho_{XE}|E) - H_{0}(\rho_{\mathbf{C}})  \nonumber \\
    	&\stackrel{(3)}{\geq} H_{\mathrm{min}}(\rho_{X \mathbf{KC}E}|\rho_{X\mathbf{K}E}) + H_{\mathrm{min}}(\rho_{XE}|E) - H_{0}(\rho_{\mathbf{C}}) \nonumber\\
    	&\stackrel{(4)}{\geq} H_{\mathrm{min}}(\rho_{X \mathbf{KC}}|\rho_{X\mathbf{K}}) + H_{\mathrm{min}}(\rho_{XE}|E) - H_{0}(\rho_{\mathbf{C}}) \label{LemmaRennereq1}
    \end{eqnarray}
    where we used: $(1)$ chain rule \cite[Section 3.1.3]{RennerThesis}, $(2)$ \autoref{Proposition} at the end of this Section, $(3)$ strong subadditivity 
    \cite[Lemma 3.1.7]{RennerThesis}, and $(4)$ Markov chain condition.\\
    To prove the general statement, for any $\varepsilon \geq 0$, let $\rho'_{XE}$ be the state $\varepsilon$-close to $\rho_{XE}$ (with respect to the purified distance)
    such that:
    \begin{equation}
    	H_{\mathrm{min}}^{\varepsilon,\,\mathrm{P}}(\rho_{X E}|E) =  H_{\mathrm{min}}(\rho'_{XE}|E)  \label{LemmaRennereq2} \,\,.
    \end{equation}
    Thanks to the definition of purified distance we can find an extension of $\rho'_{XE}$, namely $\rho'_{X\mathbf{K}E}$, such that it is still $\varepsilon$-close to
    $\rho_{X\mathbf{K}E}=\Tr_{\mathbf{C}}[\rho_{X\mathbf{KC}E}]$ \cite[Corollary 9]{purifieddistanceAdvantage}. We can assume, without loss of generality, that 
    $\rho'_{X\mathbf{K}E}$ is classical on $X$ and $\mathbf{K}$ and that $\rho'_{X\mathbf{K}}$ has support contained in the support of $\rho_{X\mathbf{K}}$\footnote{It is
    always possible to turn subsystems into classical ones by applying a CPTP map that projects onto the elements of a fixed ``classical'' basis. Note that such a map
    cannot increase the distance between states.}. Furthermore, let $\mathcal{R}_{X\mathbf{K}\rightarrow X\mathbf{KC}}$ be the CPTP recovery map that 
    recovers $\mathbf{C}$ from $(X,\mathbf{K})$, i.e.: $\rho_{X\mathbf{KC}}=\mathcal{R}_{X\mathbf{K}\rightarrow X\mathbf{KC}} (\rho_{X\mathbf{K}})$. 
    Since $X$, $\mathbf{K}$ and $\mathbf{C}$ are classical, this map can be chosen to be of the form:
    \begin{equation}
    	\fl\mathcal{R}_{X\mathbf{K}\rightarrow X\mathbf{KC}}:\,\, Q_{X\mathbf{K}}\mapsto\sum_{x,\mathbf{k},\mathbf{c}} 
    	P_{\mathbf{C}|X\mathbf{K}} (\mathbf{c}|x,\mathbf{k}) \bra{x}\bra{\mathbf{k}}Q_{X\mathbf{K}}\ket{x}\ket{\mathbf{k}}\,\, \ket{x}\bra{x}\otimes
    	\ket{\mathbf{k}}\bra{\mathbf{k}}\otimes\ket{\mathbf{c}}\bra{\mathbf{c}}  \nonumber 
    \end{equation}
    where $P_{\mathbf{C}|X\mathbf{K}}$ is the conditional probability distribution defined by the EC protocol which led to the given state $\rho_{X\mathbf{KC}E}$.
    According to the definition of min-entropy (\ref{min-entropy-rel}), for any $Q_{X\mathbf{K}}$ that is classical on $X$ and $\mathbf{K}$ 
    we have that:
    \begin{equation}
    	H_{\mathrm{min}}(\mathcal{R}_{X\mathbf{K}\rightarrow X\mathbf{KC}}(Q_{X\mathbf{K}})|Q_{X\mathbf{K}}) = -\log_2 \lambda  \label{LemmaRennereq3}
    \end{equation}
    where $\lambda$ is the minimum real number that satisfies the inequality:
    \begin{eqnarray}
    	\fl\lambda \,\mathrm{id}_{\mathbf{C}}\otimes Q_{X\mathbf{K}} - \sum_{x,\mathbf{k},\mathbf{c}} 
    	P_{\mathbf{C}|X\mathbf{K}} (\mathbf{c}|x,\mathbf{k}) \bra{x}\bra{\mathbf{k}}Q_{X\mathbf{K}}\ket{x}\ket{\mathbf{k}} \ket{x}\bra{x}\otimes
    	\ket{\mathbf{k}}\bra{\mathbf{k}}\otimes\ket{\mathbf{c}}\bra{\mathbf{c}} \geq 0 \,\,, \nonumber
    \end{eqnarray} 
    or equivalently:
    \begin{equation}
    	\lambda - P_{\mathbf{C}|X\mathbf{K}} (\mathbf{c}|x,\mathbf{k}) \geq 0 \quad\forall\, x,\mathbf{k},\mathbf{c} :
    	\bra{x}\bra{\mathbf{k}}Q_{X\mathbf{K}}\ket{x}\ket{\mathbf{k}} >0  \,\,.  \label{LemmaRennereq4}
    \end{equation}
    The minimum $\lambda$ satisfying (\ref{LemmaRennereq4}) is the maximum eigenvalue of the non-normalized state $\sum_{\mathbf{c}} P_{\mathbf{C}|X\mathbf{K}}
    (\mathbf{c}|x,\mathbf{k}) \ket{\mathbf{c}} \bra{\mathbf{c}}$, further maximized over $x$ and $\mathbf{k}$. Thus from \cite[Remark 3.1.3]{RennerThesis}   
    combined with (\ref{LemmaRennereq3}) we get: 
    \begin{equation}
     \fl H_{\mathrm{min}}(\mathcal{R}_{X\mathbf{K}\rightarrow X\mathbf{KC}}(Q_{X\mathbf{K}})|Q_{X\mathbf{K}}) =
     \inf_{ x,\mathbf{k}\,:\,\bra{x}\bra{\mathbf{k}}Q_{X\mathbf{K}}\ket{x}\ket{\mathbf{k}} >0} 
     H_{\mathrm{min}}\bigg(\sum_{\mathbf{c}} P_{\mathbf{C}|X\mathbf{K}} (\mathbf{c}|x,\mathbf{k}) \ket{\mathbf{c}} \bra{\mathbf{c}} \bigg) \label{LemmaRennereq5}.
    \end{equation}
    Because $\rho_{X\mathbf{KC}E}$ satisfies the Markov condition $\mathbf{C}\leftrightarrow(X,\mathbf{K})\leftrightarrow E$, we have:
    \begin{equation}
    	\rho_{X\mathbf{KC}E} = (\mathcal{R}_{X\mathbf{K}\rightarrow X\mathbf{KC}} \otimes \mathrm{id}_E)(\rho_{X\mathbf{K}E}) \,\,. \nonumber
    \end{equation}
    Therefore, defining:
    \begin{equation}
    \rho'_{X\mathbf{KC}E} = (\mathcal{R}_{X\mathbf{K}\rightarrow X\mathbf{KC}} \otimes \mathrm{id}_E)(\rho'_{X\mathbf{K}E})  \nonumber
    \end{equation}
    and using the fact that CPTP maps cannot increase the distance between states, $\rho'_{X\mathbf{KC}E}$ is $\varepsilon$-close to $\rho_{X\mathbf{KC}E}$, so that:
    \begin{equation}
    	H_{\mathrm{min}}^{\varepsilon,\, \mathrm{P}}(\rho_{X \mathbf{C}E}|\mathbf{C}E) \geq H_{\mathrm{min}}(\rho'_{X \mathbf{C}E}|\mathbf{C}E)\,\,. \label{LemmaRennereq6}
    \end{equation}
    Furthermore, since $\mathrm{supp}(\rho'_{X\mathbf{K}})\subseteq \mathrm{supp}(\rho_{X\mathbf{K}})$, the action of the recovery map is such that 
    $\mathrm{supp}(\rho'_{\mathbf{C}})\subseteq \mathrm{supp}(\rho_{\mathbf{C}})$, and hence by \cite[Remark 3.1.3]{RennerThesis} it holds:
    \begin{equation}
    	H_0 (\rho_{\mathbf{C}}) \geq H_0 (\rho'_{\mathbf{C}})  \label{LemmaRennereq7} \,\,.
    \end{equation}
    Note also that, because of (\ref{LemmaRennereq5}), the min-entropy of $\mathbf{C}$ conditioned on $X$ and $\mathbf{K}$ of any classical state $Q_{X\mathbf{K}}$ 
    only depends on the recovery map $\mathcal{R}_{X\mathbf{K}\rightarrow X\mathbf{KC}}$ and on the support of $Q_{X\mathbf{K}}$. Since the support of 
    $\rho'_{X\mathbf{K}}$ is contained in the support of $\rho_{X\mathbf{K}}$, we have:
    \begin{equation}
    	H_{\mathrm{min}}(\rho_{X\mathbf{KC}}|\rho_{X\mathbf{K}}) \leq H_{\mathrm{min}}(\rho'_{X\mathbf{KC}}|\rho'_{X\mathbf{K}}) \label{LemmaRennereq8} \,\,.
    \end{equation}
    Since $\rho'_{X\mathbf{KC}E}$ by construction satisfies the Markov chain condition, inequality (\ref{LemmaRennereq1}) also holds for this operator, i.e.:
    \begin{equation}
    	H_{\mathrm{min}}(\rho'_{X \mathbf{C}E}|\mathbf{C}E) \geq 
    	H_{\mathrm{min}}(\rho'_{X \mathbf{KC}}|\rho'_{X\mathbf{K}}) + H_{\mathrm{min}}(\rho'_{XE}|E) - H_{0}(\rho'_{\mathbf{C}}) \,\,. \label{LemmaRennereq9}
    \end{equation}
    Combining (\ref{LemmaRennereq9}), (\ref{LemmaRennereq2}), (\ref{LemmaRennereq6}), (\ref{LemmaRennereq7}) and (\ref{LemmaRennereq8}) yields the claim.\hfill\opensquare

    \begin{proposition} \label{Proposition}
    	For any density operator $\rho_{ABC}$:
    	\begin{equation}
    		H_{\mathrm{min}}(\rho_{ABC}|C) \geq H_{\mathrm{min}}(\rho_{ABC}|\rho_{BC}) + H_{\mathrm{min}}(\rho_{BC}|C)   \,\,.\nonumber
    	\end{equation}
    \end{proposition}
    \textit{Proof}. By definition of min-entropy (\ref{min-entropy}), there exists a density operator $\sigma_C$ such that:
    \begin{equation}
    	\rho_{BC}  \leq 2^{-H_{\mathrm{min}}(\rho_{BC}|C)} \mathrm{id}_B \otimes \sigma_C  \,.
    \end{equation}
    We thus have:
    \begin{equation}
    	\fl \rho_{ABC} \leq 2^{-H_{\mathrm{min}}(\rho_{ABC}|\rho_{BC})} \mathrm{id}_A\otimes \rho_{BC} \leq
    	2^{-H_{\mathrm{min}}(\rho_{ABC}|\rho_{BC})-H_{\mathrm{min}}(\rho_{BC}|C)} \mathrm{id}_{AB} \otimes \sigma_C
    \end{equation}
    which implies the claim. \hfill\opensquare

\end{document}